\documentclass[letterpaper]{article} 
\usepackage{aaai24}  
\usepackage{times}  
\usepackage{helvet}  
\usepackage{courier}  
\usepackage[hyphens]{url}  
\usepackage{graphicx} 
\usepackage{natbib}  
\usepackage{caption} 
\usepackage{booktabs}
\usepackage{amssymb} 

\usepackage{multirow}
\usepackage{filecontents}
\usepackage{adjustbox}
\usepackage{pgfcalendar}
\usepackage{pgfplots}
\usepackage{pgfplotstable}

\pgfplotsset{compat=1.18}
\usepackage{caption}
\usepackage{subcaption}
\usepackage{pgfplots}
\pgfplotsset{compat=1.18}
\usepackage{caption}
\usepackage{subcaption}
\usepackage{pgfplotstable}

\usepackage{pgfplotstable}
\usepackage{adjustbox}
\usepackage{tikz}
\usetikzlibrary{shapes.geometric, calc, intersections}

\usepackage{algorithm}

\usepackage{newfloat}
\usepackage{listings}
\usepackage{pgfplots}
\usepgfplotslibrary{dateplot}
\usepackage{adjustbox}

\usepackage{pgfplotstable} 

\definecolor{mypurple}{RGB}{204, 204, 255 }
\definecolor{thisblue}{RGB}{100, 149, 237}
\definecolor{thisgreen}{RGB}{159, 226, 191}
\definecolor{darkgray}{RGB}{105,105,105}
\definecolor{thisorange}{RGB}{255, 127, 80}

\DeclareCaptionStyle{ruled}{labelfont=normalfont,labelsep=colon,strut=off} 
\floatstyle{ruled}
\newfloat{listing}{tb}{lst}{}
\floatname{listing}{Listing}
\usepackage{filecontents}

\usepgfplotslibrary{groupplots}

\author{
    Patrick Gerard\textsuperscript{\rm 1}, 
    William Theisen\textsuperscript{\rm 2}, 
    Tim Weninger\textsuperscript{\rm 2},
    Kristina Lerman\textsuperscript{\rm 1}
}
\affiliations{
    \textsuperscript{\rm 1}Information Sciences Institute, University of Southern California, Marina del Rey, California, United States\\
    \textsuperscript{\rm 2}University of Notre Dame, Notre Dame, Indiana, United States\\
    pgerard@isi.edu, wtheisen@nd.edu,  tweninger@nd.edu,
    lerman@isi.edu
}

\title{Introducing and Exploring othering Across Domains: A Novel Framework Leveraging Large Language Models}
\title{Cross-Domain Exploration of Otherism with Large Language Models and Rapid Domain Adaptation}
\title{Othering Language in Russia-Ukraine War Propaganda \textcolor{red}{[update]}}
\title{Quantifying Online Expressions of Othering in Intergroup Conflict}
\title{Online Expressions of Othering and Intergroup Conflict among Russia-Ukraine War Bloggers}
\title{A Framework for Analyzing Othering and Intergroup Conflict: Online Expressions Among Russia-Ukraine War Bloggers}

\title{Decoding Othering Language: A Study of Intergroup Conflict among Russia-Ukraine War Bloggers}    
\title{Decoding the Language of Othering: A Study of Intergroup Conflict \\during Russia-Ukraine War}    
\title{The Language of Intergroup Conflict: \\Decoding Othering Among Russia-Ukraine War Bloggers}    
\title{Fear and Loathing on the Frontline: Decoding the Language of Othering by Russia-Ukraine War Bloggers}

\usepackage{bibentry}
\usepackage{algpseudocode}

\usetikzlibrary{plotmarks}

\pgfplotsset{width=6cm,compat=1.15}

\begin{document}

\maketitle

\begin{abstract}

Othering, the act of portraying outgroups as fundamentally different from the ingroup, often escalates into framing them as existential threats—fueling intergroup conflict and justifying exclusion and violence. These dynamics are alarmingly pervasive, spanning from the extreme historical examples of genocides against minorities in Germany and Rwanda to the ongoing violence and rhetoric targeting migrants in the US and Europe. While concepts like hate speech and fear speech have been explored in existing literature, they capture only part of this broader and more nuanced dynamic which can often be harder to detect, particularly in online speech and propaganda. To address this challenge, we introduce a novel computational framework that leverages large language models (LLMs) to quantify othering across diverse contexts, extending beyond traditional linguistic indicators of hostility. Applying the model to real-world data from Telegram war bloggers and political discussions on Gab reveals how othering escalates during conflicts, interacts with moral language, and garners significant attention, particularly during periods of crisis. Our framework, designed to offer deeper insights into othering dynamics, combines with a rapid adaptation process to provide essential tools for mitigating othering's adverse impacts on social cohesion.\footnote{Code to reproduce our models and findings available at \url{https://anonymous.4open.science/r/othering_language-68FF/}}


\end{abstract}



\definecolor{red}{RGB}{ 204, 0, 51 }
\definecolor{blue}{RGB}{  0, 102, 255 }

\section{Introduction}
In times of crisis, people turn to social media for information to help them make sense of events. Consequently, social media platforms can significantly influence individual perceptions and understanding of reality. This dynamic creates a strong incentive for various actors to manipulate public perceptions through online messaging and propaganda.
Tactics that frame certain groups as separate and inherently dangerous are particularly effective at evoking strong emotional responses~\cite{ saha2021short}, often contributing to radicalization and intergroup conflict~\cite{cervone2021language}. This rhetoric spans from explicit hate speech and dehumanization~\cite{buyse2014words, kennedy2023moral} to subtler forms of fear speech~\cite{buyse2014words, saha2021short, schulze2023fear, saha2023rise}.

While previous research studied some manifestations of this dynamic, like hate speech~\cite{basile2019semeval, waseem2016hateful, kennedy2018gab} and fear speech~\cite{saha2021short, saha2023rise}, these represent just pieces of the broader social process of \textit{othering}. Othering involves the systematic construction of an outgroup---the other---as fundamentally different and ultimately threatening to the ingroup. The goal of othering is to exclude and marginalize certain groups based on arbitrary characteristics like race, ethnicity or religion, ultimately establishing a self-sustaining group identity-based `us-versus-them' mentality.




Historically, othering has been used to justify extreme measures and perpetuate cycles of violence and exclusion. Across various societies and periods, outgroups have been constructed as threats to the social, cultural, or political stability of the ingroup, legitimizing repression and violence. In Nazi Germany, discursive strategies mobilized hatred against Jews, Romani people, and homosexuals, reducing them to targets of mass extermination based solely on their group membership \cite{reicher2008making}. Similarly, Stalinist terror was framed as a moral battle against an evil `other', such as ``wreckers, diversionists, and spies,'' with repression framed as necessary for the preservation of the state \cite{gerwarth2007dictators}. In India, Hindutva organizations invoked narratives of Hindu tolerance to justify violence against Muslims, blaming conflicts on supposed Muslim intolerance \cite{kaur2005performative}. 
In recent years, social media has increasingly been used to spread narratives depicting immigrants and ethnic minorities as threats to national and cultural values~\cite{nortio2021nightmare}. In one particularly egregious example, Facebook was used to incite deadly riots and genocidal behavior in Myanmar against the Rohingya minority~\cite{yue2019weaponization}.
Ultimately, this process—--driven by persistent themes of intergroup conflict and perceived threats---endures across communities and time, adapting to the evolving needs of the ingroup, serving as a self-sustaining force of exclusion. As such, tracking and quantifying this social dynamic on online platforms is crucial. 
While existing research has extensively explored explicit manifestations of intergroup conflict like hate speech and fear speech~\cite{ cervone2021language, saha2021short}, these expressions are only part of the broader process of othering. Our work explores the subtler, lesser-understood mechanisms that sustain these expressions, shedding light on their nature and their 
interactions with moral language and attention mechanisms.

We ground our analysis in sociological theory, defining `othering' as the  process of depicting a group as fundamentally different from one's own~\cite{reicher2008making, cikara2015intergroup, jetten1997strength}, this group self-talk is often characterized by the positive portrayal of one's own group (i.e., the ingroup) and the negative portrayal of the other group (i.e., the outgroup). This social dynamic marginalizes, excludes, or discriminates against the outgroup based on arbitrary or perceived differences, such as race, religion, or ethnicity, reinforcing social hierarchies and justifying unequal treatment~\cite{reicher2008making, duckitt2003prejudice}. We introduce a taxonomy of othering language and show that it subsumes and extends commonly used text indicators of intergroup conflict.

Additionally, we introduce an innovative computational framework that leverages large language models (LLMs) as classifiers and enables their rapid adaptation to new domains.
After validating a model of othering, we use it to explore the language of intergroup conflict in real-world scenarios. We analyze a corpus of messages posted on Telegram by Russian and Ukrainian war bloggers during the ongoing war between Russia and Ukraine, as well as a corpus of messages posted on the social media platform Gab. 
We explore the following research questions:
\begin{enumerate}
    \item How does the use of othering by Russian and Ukrainian war bloggers on Telegram change over the course of the war?
    \item How do moral language and othering interact, especially in the expressions of intergroup conflict? 
    \item How does constructing and reinforcing the image of a target group as the `other' affect social attention?
    \item Does othering intensify during times of crisis, and in what ways are these behaviors more strongly rewarded?
\end{enumerate}
Our analysis reveals the amplification of othering language in polarized online environments and its tendency to attract attention, especially during crises. While we find that othering language is often moralized across groups, its asymmetrical use by Russian war bloggers highlights its distinct utility in propaganda. By exploring these dimensions, we demonstrate how othering underpins more overt expressions like fear speech, hate speech, and exclusionary practices, offering crucial insights for developing strategies to counteract othering and mitigate its impact on social cohesion.


\section{Related Work and Background}

\paragraph{The Language of Intergroup Conflict}
Intergroup conflict often drives violence by framing outgroups as existential threats to the ingroup. From Nazi Germany to modern online spaces, fear-mongering and hateful rhetoric radicalize populations by portraying outgroups as dangerous and immoral, justifying hostility and violence~\cite{greipl2022online, reicher2008making}.

Such conflict escalates during crises—pandemics, financial collapses, or political upheaval—when fears are externalized and outgroups are scapegoated for societal problems. Economic hardship heightens competition and prejudice, as seen during the Great Depression. Similarly, the 2015 European Refugee Crisis saw the rise of exclusionary rhetoric, driven by political and social tensions~\cite{pettersson2017pray}. Misperceptions of outgroup hostility further exacerbate tensions, with groups mistakenly believing the other supports violence, as seen in the 2021 Israeli-Palestinian conflict. However, corrective interventions have shown promise in reducing these tensions~\cite{nir2023kill}.

A key psychological mechanism driving intergroup conflict is the perception of outgroup threat--—the belief that outgroups endanger the ingroup’s identity or very existence. This dynamic was particularly evident during the COVID-19 pandemic, where heightened awareness of mortality intensified xenophobia~\cite{esses2021xenophobia}.

Ultimately, these often-manufactured perceptions of threat can escalate intergroup conflict, leading to scapegoating that legitimizes violence and perpetuates hatred. This cycle frequently results in real-world violence, reinforcing instability and deepening social divisions~\cite{fink2018dangerous, warofka2018independent}.

Computational scientists often operationalize the language of intergroup conflict through two key concepts: hate speech and fear speech. Hate speech refers to expressions intended to insult, degrade, or incite hostility toward a group based on attributes like race, religion, or gender~\cite{mathew2021hatexplain}. It promotes explicit hostility typically through vilification and dehumanization~\cite{basile2019semeval, waseem2016hateful, kennedy2018gab}. In contrast, fear speech invokes existential fear, portraying the target group as a fundamental threat to the ingroup's survival, culture, or identity~\cite{buyse2014words, saha2021short, saha2023rise}. Both forms of speech reinforce an 'us-versus-them' mentality, either by inciting hostility or by instilling fear, emphasizing the danger the outgroup poses to the ingroup’s way of life. Together, they contribute to the broader process of othering, where groups are marginalized or excluded based on perceived differences~\cite{reicher2008making, cikara2015intergroup, jetten1997strength}."


\paragraph{Social Mechanisms of Othering}
Othering language extends beyond specific expressions like hate speech and fear speech; it encompasses the broader sociological process of constructing an outgroup and excluding individuals based on race, religion, or ethnicity. This process has been observed throughout history, from the Holocaust to contemporary political discourse on immigration~\cite{reicher2008making}. Understanding these mechanisms is essential for mitigating their effects and preventing intergroup conflict.

\begin{figure} 
\hspace{-6mm}
\includegraphics[width=0.53\textwidth]{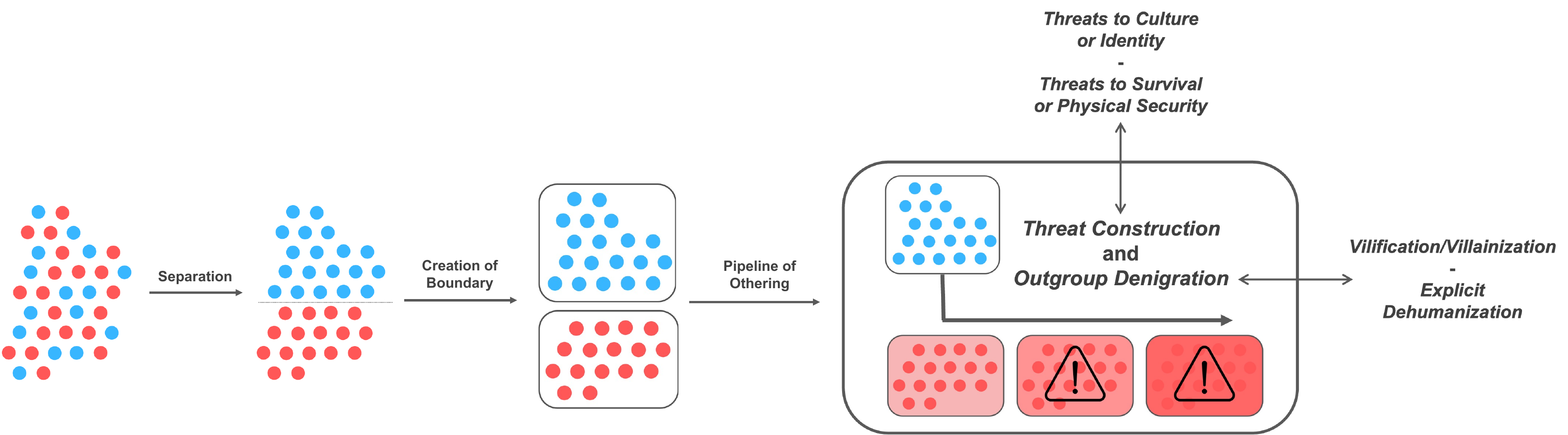} 

\vspace{-1mm} 
\caption{Conceptualization of the  othering process. Othering starts with the separation of ingroup and outgroup members, the creation of symbolic boundaries, and the subsequent pipeline of othering. Through the construction and affirmation of perceived threats, the outgroup is increasingly framed as a threat.} 
\label{fig:otherism_diagram}
\vspace{-2mm} 
\end{figure}

We base our understanding of othering on Reicher’s model~\cite{reicher2008making}, which conceptualizes hate as emerging from a continuous process of othering. As illustrated in Figure~\ref{fig:otherism_diagram}, extreme violence and genocide are driven by a distorted perception of group identity, where individuals are targeted solely for belonging to an outgroup~\cite{reicher2008making, kaur2005performative}. In this process, even disavowing one’s group identity offers no protection, as othering reduces individuals to their perceived group membership, overriding personal actions or beliefs~\cite{reicher2008making, duckitt2003prejudice}.

The model outlines five key steps in the development of collective hate: 
(i) creating a cohesive ingroup, 
(ii) excluding specific populations, 
(iii) framing the outgroup as a threat to the ingroup’s existence, 
(iv) portraying the ingroup as virtuous, and 
(v) celebrating the outgroup’s destruction as a defense of ingroup virtue \cite{reicher2008making}. Central to this model is the formation of both ingroup and outgroup boundaries, which together play a critical role in fostering hostility and ultimately perceiving the outgroup as a threat~\cite{reicher2008making}.
This dynamic is evident in ideologies like Nazism, where outgroups were framed as existential threats, justifying violence as moral and necessary for 'cleansing' and protecting the ingroup’s values~\cite{kennedy2023moral, koonz2003nazi}. Similarly, contemporary threats to the ingroup—whether economic or cultural—are often constructed to legitimize hostility. These identity narratives are not static but actively constructed, with the perceived danger of outgroups adapting to the fears of the ingroup~\cite{reicher2008making}.

The intersection of othering and moralized language is crucial for understanding how narratives of intergroup conflict are constructed. Moral language often serves as a powerful tool for legitimizing exclusion and violence~\cite{fiske2014virtuous, kennedy2023moral}. According to the \textit{Moralized Threat Hypothesis}, extreme expressions of prejudice are frequently driven by the belief that the outgroup has committed a moral transgression, making violence not only justified but also morally righteous~\cite{hoover2021investigating}. Consequently, we expect to find strong correlations between othering language—where a group is systematically framed as an enemy—and moral language.

Moreover, othering’s role in capturing social attention demands further study. Narratives built around othering often attract disproportionate attention in online spaces, as the portrayal of an outgroup as a threat captures the audience’s focus \cite{saha2023rise}. This heightened attention amplifies the spread and impact of these narratives, particularly during times of crisis or polarization. Understanding this dynamic is essential for grasping how othering influences public discourse and fosters division

\paragraph{LLMs and In-Context Learning}
In-context learning enables LLMs to perform tasks by leveraging a few input-label pairs, or demonstrations, without requiring gradient updates, and has often outperformed zero-shot learning across numerous tasks~\citep{zhao2021calibrate}. However, its success is influenced by factors such as task complexity, the quality of provided examples, and the number of demonstrations. Our approach, which uses the \textit{system prompt} to guide an LLM fine-tuned on one domain to adapt to another, offers a similar but distinct method for applying LLMs to unseen and untrained data.
\section{Methods}
\subsection{Data}

\subsubsection{Russia-Ukraine war bloggers}
\label{sec:telegram}
We use posts collected from Russian-oriented and Ukrainian-oriented Telegram channels~\cite{theisen2022motif}, spanning from October 2015 to August 2023. This dataset, which was gathered using both an expert-curated list and snowballing methods, includes 989 channels and over 9.67 million posts, primarily in Ukrainian and Russian. This data is released alongside this paper to support our findings and compel further research.

\begin{figure} 
\includegraphics[width=0.5\textwidth]{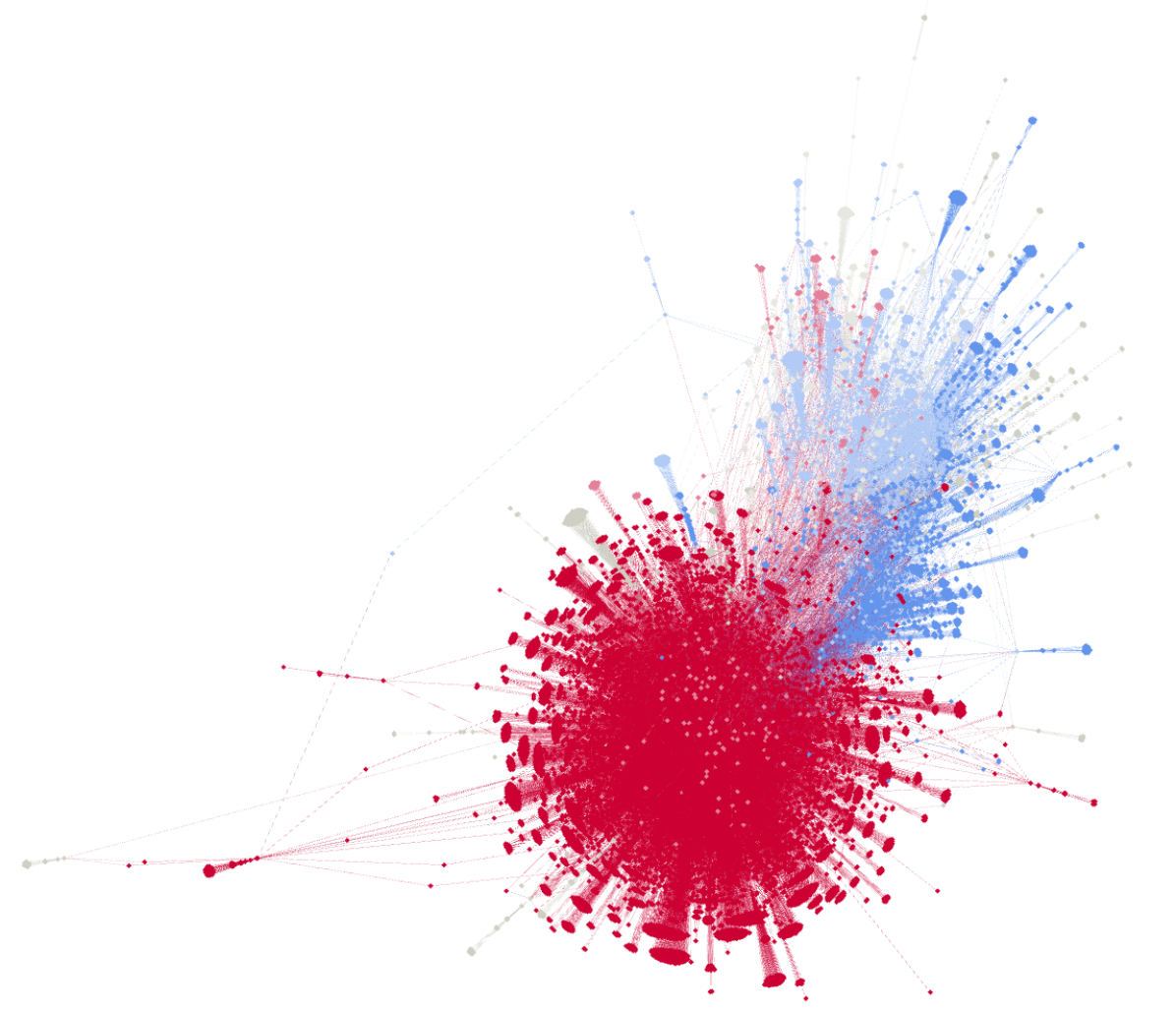} 
\caption{        Visualization of a co-reference network based on message content and channel bios. 
        Nodes are colored according to the stance indicated by their activity:
        \textcolor{red}{Pro-Russian} nodes are in red,
        \textcolor{blue}{Pro-Ukrainian} nodes are in blue, and nodes not strongly affiliated to either nationality are in grey.
        This network was constructed by analyzing messages and bio information from various Telegram channels.
} 
\label{fig:warblogger_network}
\end{figure}

\noindent
\textit{Labeling Telegram Channels:}
Telegram, a messaging app supporting private/public group interactions and one-way broadcasts via channels, has become a stronghold of a `free' Internet in Russia. Evading bans affecting other major platforms like like Facebook and TikTok since 2020, Telegram has emerged as a key platform for military (war) bloggers and a primary source of news for both Ukrainians and Russians during the Russia-Ukraine war~\cite{oleinik2024telegram}.
We construct a network of Telegram channels within the war bloggers corpus, where a directed link with weight $w$ connects channel A to channel B if A references or forwards B’s posts $w$ times during the period. 
The network, Figure \ref{fig:warblogger_network}, shows the channels and connections between them. The global structure of information sharing shows roughly two  clusters, as expected of  groups in conflict.

We manually labeled 100 random channels as ``pro-Russia', `pro-Ukraine,' or `Other' based on their bios and recent posts, then used these as seeds for a label propagation algorithm~\cite{garza2019community} to categorize the network. To validate, we reviewed 100 randomly selected channels from each group. The final dataset includes 243 pro-Ukrainian channels (~4.2 million posts) and 325 pro-Russian channels (~4.4 million posts) from October 2015 to August 2023, though messages prior to late 2021 are sparse.

\subsubsection{Gab Corpus}
Introduced in prior work by Saha et al. \cite{saha2023rise}, this corpus contains 9,441 text posts from the popular `alt-tech' social media platform Gab \cite{dehghan2022politicization}. Gab, which is popular with political conservatives in US, hosts discussions often revolving around issues of race, immigration, and national identity. Despite the absence of direct physical conflict, the rhetoric on Gab is steeped in othering narratives, making it a platform of choice for studying hate speech and fear speech~\cite{kennedy2018gab, saha2023rise}. Each post was manually classified by humans annotators into one or more categories: `normal,' `fear speech,' or `hate speech,' with some posts receiving multiple labels. We detail the definitions used for fear speech and hate speech in a later section on othering language's relationship to expressions of intergroup conflict. Overall, 44.8\% of the posts are labeled as normal, 19.7\% as fear speech, and 42.4\% as hate speech

\subsection{A Model of Othering}
We develop a flexible, LLM-based model to recognize othering language in text and show how it can be rapidly adapted to new domains.

\subsubsection{Taxonomy of Othering}
Othering is a group self-talk process that helps shape group's conceptualization of itself as good and virtuous and the other group as inherently evil and dangerous. When talking about itself, i.e., the ingroup, the group uses fear-laden speech~\cite{lerman2024affective}, which serves to bind the group together, often in response to a perceived threat from the outgroup. When talking about the other, i.e., the outgroup, othering manifests itself through animosity and hostility~\cite{stephan2015intergroup, joffe1999risk}. 
To capture the various dimensions of othering, we define  four categories of language linked to the othering process and provide translated examples from Russian war bloggers. The first two categories address perceived threats to the ingroup, while the latter two focus on the demonization of outgroups.
\begin{description}
    \item[{Threats to Culture or Identity}] arise when the outgroup is framed as a danger to the ingroup’s cultural or social survival, challenging its values, language or traditions~\cite{wohl2010perceiving, reicher2008making, stephan2015intergroup, joffe1999risk}. Example post: \textit{``The erosion of the Russian language in Ukrainian schools: Ukrainian policymakers pushing to erase the Russian tongue risk severing the threads that weave together our history.''}
    
    \item[{Threats to Survival or Physical Security}] involve portraying the outgroup as an existential menace to the ingroup’s physical well-being, thereby justifying preemptive hostility \cite{wohl2010perceiving, reicher2008making, stephan2015intergroup, joffe1999risk}. Example post: \textit{``Zelensky's regime has accumulated 30 tons of plutonium and 40 tons of enriched uranium at the Zaporizhia NPP [...] the regime really is on the verge of creating its own nuclear bomb! And hundreds of `dirty' bombs can be made from such a quantity of radioactive material!''}
    \item[{Vilification/Villainization}] 
    casts the outgroup as inherently evil or immoral, which in turn validates resistance and aggression \cite{joffe1999risk, reicher2008making, stephan2015intergroup}. Example post: \textit{``Because these Ukronazi girls can fight only by hiding behind hostages. All their courage went down the drain in chants and slogans like `hang the Muscovite.' But when the Russians came, they shit themselves, just like their Bandera.''}
    \item[{Explicit Dehumanization}] represents the most extreme form of othering, where the outgroup is compared to animals, objects or spirits, paving the way for extreme violence \cite{joffe1999risk, reicher2008making, stephan2015intergroup}. Example post: \textit{``These are zombies, who may have been brothers before, but over the past 8 years, from the bite of Nazism and Banderization, they have turned into non-humans. That is why our army calls on all brothers to lay down their arms, so that we can distinguish a brother from an infected zombie, who can only bite and infect.''}




   
\end{description}

These classes are integral to understanding how outgroups are systematically constructed, legitimizing their exclusion and violence.

\subsubsection{Artificial Annotator Alignment Process}
\begin{figure} 
\includegraphics[width=0.45\textwidth]{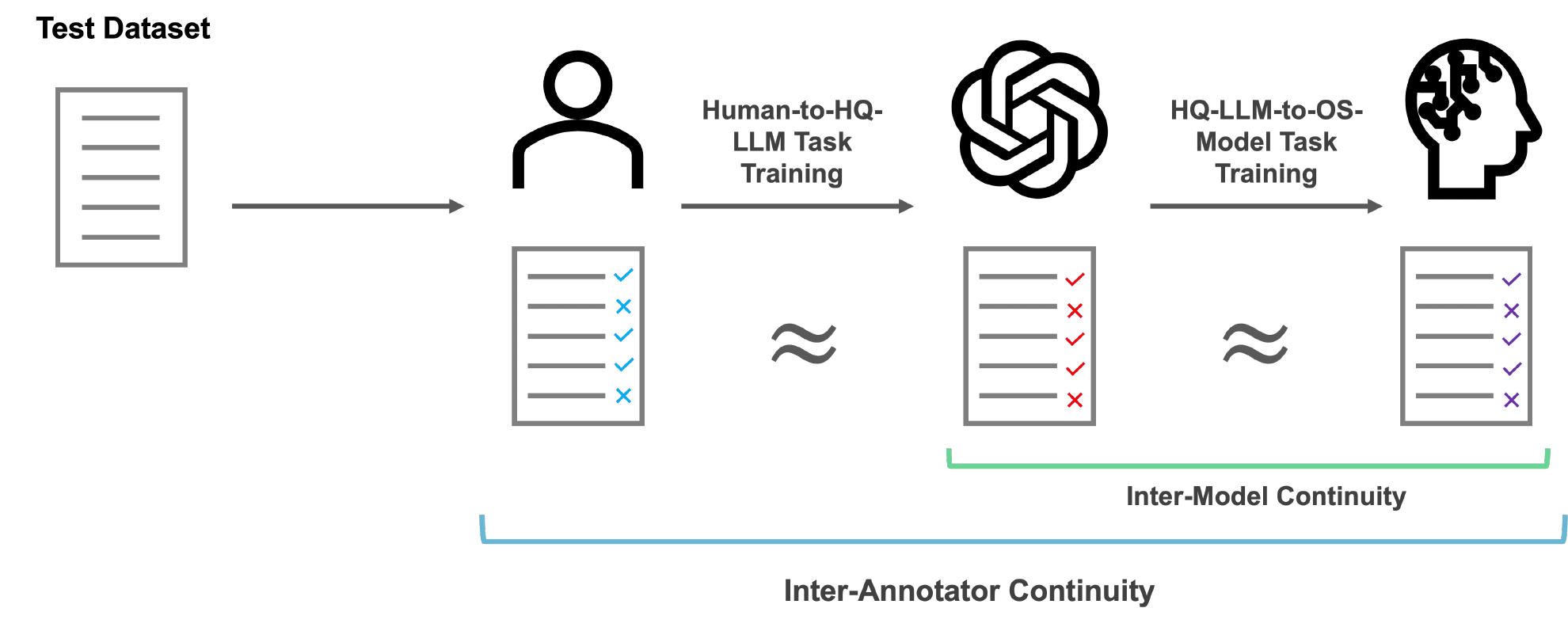} 
\vspace{-2mm} 
\caption{Artificial Annotator Alignment process. Human-annotated data is first used to train a high-quality LLM. The HQ-LLM's annotations are compared with human annotations for alignment, and then the HQ-LLM is used to annotate a larger dataset. Finally, an open-source model is trained using the HQ-LLM-annotated data, optimizing both effectiveness and cost-efficiency.} 
\label{fig:knowledge_distillation}
\vspace{-4mm} 
\end{figure}

We train a classifier to recognize othering language, using an ``Artificial Annotator Alignment'' process inspired by knowledge distillation~\cite{gou2021knowledge} to efficiently train the model and adapt it to new domains. Our alignment process, shown in Figure \ref{fig:knowledge_distillation}, essentially ``trains'' LLMs as annotators by requiring them to be consistent with human annotators. Fist, human annotators label a small subset of the data. This subset is then annotated by a ``high-quality'' (HQ) LLM like ChatGPT-4. If the HQ-LLM's annotations closely align with those of the human annotators, we proceed to have the HQ-LLM annotate more data to effectively train an open-source LLM (OS-LLM), e.g., Llama3, as an annotator. This strategy is driven by the goal of maximizing effectiveness and cost: utilizing a combination of an HQ-LLM and OS-LLM to avoid the prohibitive costs of large-scale annotation.

This approach is guided by two core principles: (1) The initial, high-quality (HQ) LLM must produce annotations that closely resemble those made by humans; (2) The open-source (OS) LLM trained on the HQ-LLM-annotated data must achieve performance on par with the HQ-LLM when evaluated against the human-annotated data (which is held-out throughout the process for evaluation). To adhere to the first principle, we assess the continuity between human annotations and the HQ-LLM using both standard machine learning metrics (accuracy, F1-score, etc.) and inter-annotator agreement metrics. This approach allows us to evaluate the HQ-LLM both as a classifier and as an artificial annotator, ensuring its consistency with human annotations. If both sets of evaluations yield optimal results, we consider the HQ-LLM a reliable proxy for human annotation. We then proceed to use the HQ-LLM to annotate a substantially larger portion of the dataset.

To adhere to the second principle, after training the target OS-LLM on the data annotated by the HQ-LLM, we  test its performance on the human-annotated dataset, once again using both sets of metrics to ensure that its classification metrics do not degrade from the HQ-LLM (ensuring inter-model continuity) and that its classifications are consistent with human annotations (ensuring inter-annotator continuity). This two-step validation ensures that the target model OS-LLM not only replicates the quality of the HQ-LLM’s annotations but also aligns with human judgment, thereby confirming its effectiveness in the domain.





\paragraph{Data Annotation}
Our analysis began with two datasets: messages from Russian war bloggers and from Ukrainian war bloggers. For each dataset, we used a combination of random post selection and keyword-based upsampling, targeting phrases and coded terms of denigration specific to each context (e.g., ``Ukronazis'' in the Russian data). 

\textit{Human Annotation:} The approach yielded 316 posts for the Russian data, which were labeled by six human annotators with overlapping annotations. A similar process was applied to the Ukrainian dataset, but with two annotators, based on the strong performance observed in the Russian workflow. Inter-annotator agreement, shown in Tables~\ref{tab:krippendorff_alpha} and \ref{tab:ukrainian_gold_set_summary} for the Russian and Ukrainian data, is consistent with benchmarks in similar studies \cite{saha2021short, saha2023rise}. Final classification counts for each domain were determined through majority vote, as summarized in Tables \ref{tab:gold_set_summary} and \ref{tab:ukrainian_gold_set_summary}.

\textit{High-Quality LLM Annotation:}
Following independent human annotations, we used ChatGPT-4o (HQ-LLM) to annotate the same examples using prompts tailored to the specific context (prompts available in our GitHub repository), outputting a dictionary for each post: ``{`Threats to Culture or Identity': 1, `Threats to Survival or Physical Security': 0, `Vilification/Villainization': 1, `Explicit Dehumanization': 0, `None': 0}'', along with an explanation. For example, ``The text describes local Nazis desecrating a historic Russian cemetery, in a way that represents a threat to cultural identity and vilifies the opposing group.'' These explanations were crucial for understanding the rationale behind each annotation and for testing our Rapid Domain Adaptation method later. The annotations were validated using  metrics, such as Cohen’s Kappa, treating ChatGPT-4o as an additional annotator. The results, detailed in tables \ref{tab:accuracy_f1_uk_gpt}, show that the ChatGPT-4o annotations were reliable and consistent across domains. In total, ChatGPT-4o annotated 20,000 posts (10,000 from each dataset) at a cost of approximately \$70 USD, significantly lower than human annotation costs while remaining consistent with human annotators.

\paragraph{Training Models}
Next, we trained three different models (OS-LLMs) on the ChatGPT-4o-annotated (HQ-LLM) data: Mistral, LLaMA3-8b-Instruct, and LLaMA2. Each model was trained separately on three datasets: Russian-only, Ukrainian-only, and a combined Russian-and-Ukrainian dataset. To ensure a fair evaluation across different domains, we withheld common test and validation sets for each dataset, preventing data leakage that could give a model an unfair advantage. Each model was trained for 5 epochs on an NVIDIA Quadro RTX 8000 GPU using a learning rate of 1e-5 and followed a 0.7:0.1:0.2 split for training, validation, and testing. The best-performing model from each run was selected based on the best F1 score on the validation set.

After training, we evaluated each model across three domains: Ukrainian-only, Russian-only, and combined Russian-and-Ukrainian. The exact performance metrics for each domain-model pair are available in our repository, with the best-performing model—LLaMA3-8b-Instruct—detailed in Table \ref{tab:accuracy_f1_ru_LLM}. Our evaluation shows that the model adheres to principle 2, exhibiting minimal degradation in ML-based metrics compared to ChatGPT-4o and remaining consistent with human annotators when assessed as an artificial annotator.

Models trained on a domain perform best on that domain but struggle to generalize to other domains. In contrast, models trained on multiple domains demonstrate better cross-domain generalization, though they underperform single-domain models within their specific domain.

\begin{table}[h]

\centering
\resizebox{0.5\textwidth}{!}{%

\begin{tabular}{lccc}
\noalign{\smallskip} 
Category & Cohen's & Accuracy & F1 Score
\\
\hline
\addlinespace[1mm]
Threats to Culture or Identity & 0.84 & 0.89 & 0.86\\ 
Threats to Survival or Physical Security & 0.74  & 0.84  & 0.80 \\ 
Vilification/Villainization & 0.81  & 0.91  & 0.89  \\ 
Explicit Dehumanization & 0.84  & 0.89  & 0.88  \\ 
None & 0.84  & 0.90  & 0.87  \\ \hline
\end{tabular}
}
\caption{Inter-Annotator Agreement and Model Performance: Cohen’s Kappa (Agreement), Accuracy, and F1 Score between Majority Vote and LLM on Russian Data.}
\label{tab:accuracy_f1_ru_LLM}
\end{table}

\begin{table}[h]
\vspace{-4mm}
\centering
\resizebox{0.5\textwidth}{!}{%
\begin{tabular}{lccc}
Category & Cohen's & Accuracy & F1 Score \\ \hline
\addlinespace[1mm]
Threats to Culture or Identity & 0.81  & 0.88  & 0.89  \\ 
Threats to Survival or Physical Security & 0.72  & 0.87  & 0.82  \\ 
Vilification/Villainization & 0.80  & 0.87  & 0.86  \\ 
Explicit Dehumanization & 0.82  & 0.89  & 0.88  \\ 
None & 0.83  & 0.88  & 0.84  \\ \hline
\end{tabular}
}
\caption{Inter-Annotator Agreement and Model Performance: Cohen’s Kappa (Agreement), Accuracy, and F1 Score between Majority Vote and LLM on Ukrainian Data.}
\label{tab:accuracy_f1_uk_LLM}
\end{table}




\vspace{-2mm}
\subsubsection{Rapid Domain Adaptation}
Adapting models to new domains presents a significant challenge in classification tasks. To study othering across multiple domains in a cost-effective manner, we test whether our models could effectively transfer knowledge from one domain to another. LLMs are well-suited for this task due to their (1) inherent complexity and (2) the pseudo-world mapping generated through extensive pretraining on vast corpora. To leverage this power, we employ two techniques that adapt the classifier to new, unseen data: system prompt steering and logit disambiguation. We name this approach Rapid Domain Adaptation (RDA).

\begin{figure} 

\includegraphics[width=0.52\textwidth]{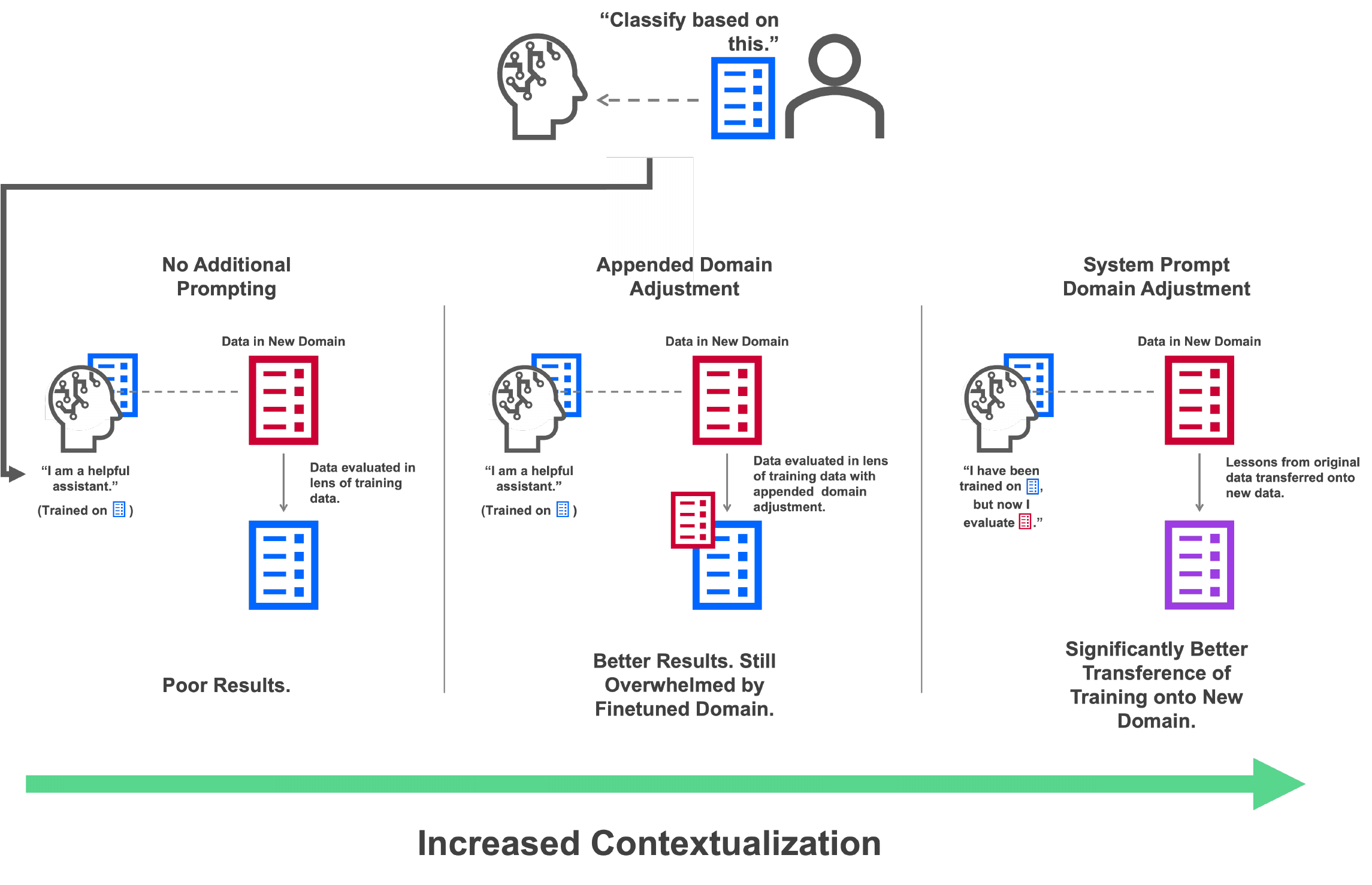} 
\vspace{-1mm} 
\caption{System Prompt Steering: demonstrates increased contextualization with the use of system prompt steering.} 
\label{fig:system_promt_engineering}
\vspace{-2mm} 
\end{figure}

\noindent\textit{System Prompt Steering:}
The first component of our RDA system involves manipulating the system prompt for the trained model. A system prompt provides the model with initial instructions, guiding how it processes and classifies incoming data. 
While in-context learning offers some benefits, simply appending context to new data only minimally improves performance. However, a well-crafted system prompt, designed to steer the model’s reasoning, led to significant improvements in new domains by explicitly framing the new domain (this logic is illustrated in Figure \ref{fig:system_promt_engineering}) in relation to the model's prior training, we achieved notable enhancements in domain adaptation.


\noindent\textit{Logit Disambiguation:}
The second component of our RDA system involves exposing the logits for each class, rather than simply assigning a binary label of 0 or 1. Logits represent the model’s confidence in its predictions, indicating the likelihood of a particular token being chosen. Since our task is to classify messages using either 1 or 0 for each class, we can expose the logits for this classification token and then using confidence thresholds for each class, fine-tuning them specifically for new domains. This approach is especially useful when the new domain features significantly different language or topics compared to the original training domain. As shown in Figure \ref{fig:logit_disambiguation}, by disambiguating the logits and adjusting confidence thresholds, we can better adapt the model to the new domain.

Overall, these components --- system prompt steering and logit disambiguation --- work together to enable rapid and reliable domain adaptation, leveraging modern LLMs to effectively handle drastic shifts in domain context.

\noindent
\textit{RDA Evaluation:}
We evaluated our RDA system across all cross-domain combinations (e.g., a model trained on Russian war blogger data performing on Ukrainian war blogger data), with detailed results available in our repository. We first compared system prompt steering to two baseline approaches: no added context and traditional in-context learning, where context is simply appended to the new prompting data. Table \ref{tab:f1_score_comparison} shows the F1 scores for different domain transitions. Figure \ref{fig:system_prompt_gain} visualizes the substantial performance gains across metrics when a model trained on Russian data applied to Gab, which contains messages posted on a different platform, in a different language (English), and in a different cultural and geopolitical context (representing the most substantial domain transition). These results demonstrate the effectiveness of our RDA system, especially in models that had no prior exposure to the new domain.
\begin{table}[h]

\centering
\resizebox{0.49\textwidth}{!}{%

\begin{tabular}{lcccc}
Prompting Type & Accuracy & F1 Score & Precision & Recall
\\
\hline
\addlinespace[1mm]
No Additional Prompt & 0.55 & 0.53 & 0.83 & 0.56 \\ 
In-Context Learning & 0.61 & 0.63 & 0.84 & 0.65 \\
System Prompt & 0.69 & 0.76 & \textbf{0.90} & 0.70 \\ 
RDA & \textbf{0.71} & \textbf{0.77} & \textbf{0.90} & \textbf{0.71} \\ 
\hline
\end{tabular}
}
\caption{Performance Comparison of Different Prompting Types: Accuracy, F1 Score, Precision, and Recall.}
\label{tab:prompting_performance}
\vspace{-2mm}
\end{table}

\definecolor{mypurple}{RGB}{204, 204, 255 }
\definecolor{thisblue}{RGB}{100, 149, 237}
\definecolor{thisgreen}{RGB}{159, 226, 191}
\definecolor{thisorange}{RGB}{255, 127, 80}

\begin{figure}[h]
    \centering
\resizebox{.4\textwidth}{!}{%
    \begin{tikzpicture}
        \begin{axis}[
            ybar,
            width=9cm,
            height=6cm,
            symbolic x coords={
                Accuracy, 
                F1 Score, 
                Precision, 
                Recall
            },
            xtick=data,
            xlabel={Metrics},
            ylabel={System Prompt \% Gain},
            ymin=0,
            ymax=40,
            bar width=15pt,
            axis line style={line width=.3pt}, 
            x tick label style={yshift=-7pt}, 
            tick style={
                major tick length=0pt, 
            },
            legend style={at={(0.5,-0.3)}, anchor=north, legend columns=-1},
            legend cell align={left},
            legend entries={System vs No Prompt, System vs In-Context},
            ymajorgrids=true,
            enlarge x limits={abs=1.00cm}, 
        ]
            \addplot[fill=thisblue!70!white]
            coordinates {
                (Accuracy, 22.58) 
                (F1 Score, 35.66) 
                (Precision, 8.09) 
                (Recall, 22.22) 
            };

            \addplot[fill=thisgreen!80!white]
            coordinates {
                (Accuracy, 12.31) 
                (F1 Score, 18.71) 
                (Precision, 6.90) 
                (Recall, 7.41) 
            };
        \end{axis}
    \end{tikzpicture}
    }
    \caption{System Prompt \% Gain Compared to No Additional Prompt and In-Context Learning Across Metrics (Accuracy, F1 Score, Precision, Recall).}
    \label{fig:system_prompt_gain}
    \vspace{-4mm}
\end{figure}
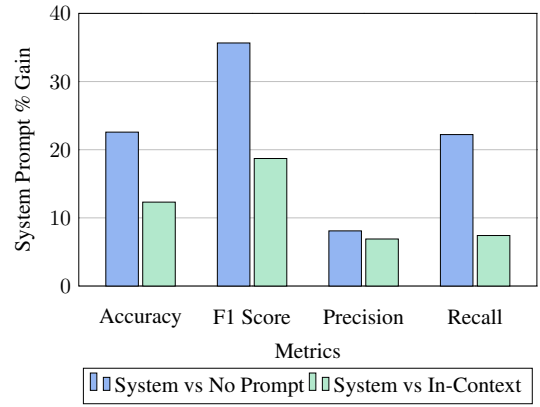

\subsection{Relationship to Language of Intergroup Conflict}
\begin{figure} 
\centering
\includegraphics[width=0.32\textwidth]{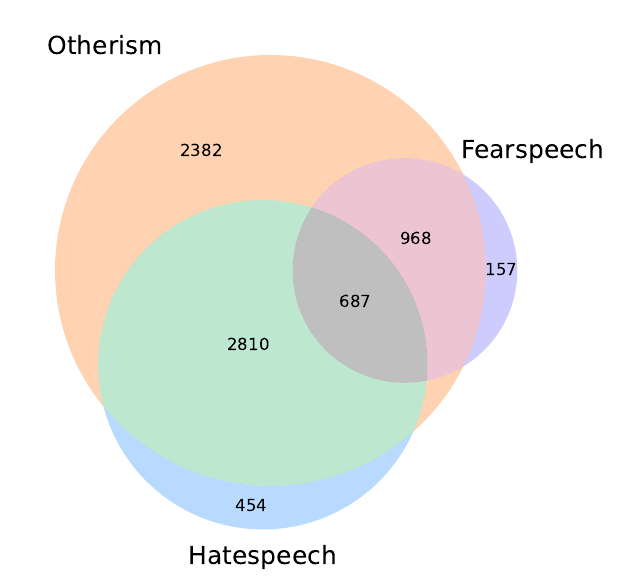} 
\vspace{-1mm} 
\caption{Venn diagram showing overlap between othering, fear speech, and hate speech in the Gab corpus. The diagram reveals that while fear speech and hate speech often co-occur with othering, many instances of othering occur without these explicit forms of conflict language.
}
\label{fig:venn_diagram}
\vspace{-2mm} 
\end{figure}

We examine the relationship between othering language and other widely studied expressions of intergroup conflict, such as fear speech and hate speech, using the annotated Gab corpus. As illustrated in Fig.~\ref{fig:venn_diagram}, fear speech and hate speech reflect expressions of othering, but they do not fully encompass it. Instead, they function as partial components of the broader process. This underscores the asymmetry suggested by our model, which posits that othering language subsumes, but is not limited to, specific expressions like fear speech and hate speech. Additionally, we consider the practical implications for content moderation by evaluating how well current toxicity classifiers detect othering language alongside fear speech and hate speech. This analysis highlights the limitations of existing classification tools in addressing the full spectrum of othering language.

\paragraph{Fear Speech}
Fear speech, defined as expressions that instill existential fear of a target group (often based on attributes such as race, religion, or gender)~\cite{buyse2014words}, is significantly associated with othering language. Specifically, the probability that fear speech contains `Vilification/Villainization' is 68.9\%, and the probability of containing `Threats to Culture or Identity' is 50.5\%. It has a weaker association with `Threats to Survival or Physical Security' at 20.3\%. These connections reflect the nature of fear speech in both evoking existential fear and subtly vilifying the outgroup, such as in messages like ``They will destroy our way of life unless we stop them.'
Moreover, fear speech shows an asymmetric relationship to othering: 88.9\% of fear speech instances involve othering, but only 24.2\% of othering messages are classified as fear speech.


\paragraph{Hate Speech}

Hate speech is language used to express hatred toward a targeted individual or group or is intended to be derogatory, to humiliate, or to insult the members of the group, based on attributes such as race, religion, or gender~\cite{mathew2021hatexplain}, and is typically the most explicit form of othering. Our analysis shows strong associations with `Vilification/Villainization' (74.1\%), `Explicit Dehumanization' (37.3\%), and `Threats to Culture or Identity' (32.3\%). These connections underscore hate speech's dual role both denigrating the outgroup and framing it as a threat.

Hate speech also demonstrates an asymmetric relationship with othering language: approximately 87.4\% of hate speech involves othering, but only 51.1\% of messages containing othering language are classified as hate speech.

\paragraph{Toxicity}
Finally, 
we analyze the relationship to toxicity. 
Using the Detoxify classifier\footnote{\url{https://github.com/unitaryai/detoxify}}, which rates text on a scale from 0 to 1 (with scores above 0.5 considered toxic), we find the following average toxicity scores: fear speech averages 0.46, while hate speech scores higher at 0.65. For othering content, excluding Explicit Dehumanization, the average toxicity is 0.42 and increases to 0.53 when Explicit Dehumanization is included (which has a notably high average toxicity of 0.81). These findings highlight the broad spectrum of toxicity within othering rhetoric and emphasize the need for more effective detection tools.

\vspace{-2mm}
\section{Results}
We label the corpus of messages posted on Telegram by Russian and Ukrainian war bloggers for othering language, focusing on the period from late 2021 to August of 2023. This period of war was characterized by high conflict and animosity between the groups. We also label a corpus of messages posted on Gab, which is often favored by far-right users within the US.

\subsection{Othering during the Russia-Ukraine War}

\begin{figure}[t!]
\vspace{-4mm}
    \hspace{-5mm}
    \begin{tabular}{c}
    \begin{tikzpicture}
    \begin{axis}[
        xlabel=(a) Russian war bloggers ,
        ylabel=Proportion of Messages,
        date coordinates in=x,
        xticklabel style={rotate=90, anchor=center, xshift=-19pt}, 
        ytick={0, 0.05, 0.1, 0.15, 0.2, 0.25}, 
        yticklabels={0, 0.05, 0.1, 0.15, 0.2, 0.25},
        yticklabel style={xshift=-5pt}, 
        xtick={2021-12-01, 2022-02-01, 2022-04-01, 2022-06-01, 2022-08-01, 2022-10-01, 2022-12-01, 2023-02-01, 2023-04-01}, 
        xticklabels={Dec '21, Feb '22, Apr '22, Jun '22, Aug '22, Oct '22, Dec '22, Feb '23, Apr '23},
        xmin=2021-12-01, xmax=2023-05-01, 
        ymin=0, ymax=0.25,
        width=\linewidth,
        height=0.6\linewidth,
        axis x line*=bottom,  
        ymajorgrids=true,
        legend style={at={(0.5,-0.57)}, anchor=north,legend columns=1, font=\small}
    ]

    \addplot[color=thisgreen!90!black, line width=1pt] table [x=date, y=Threats to Culture or Identity, col sep=comma] {proportions_over_time_ru.csv};

    \addplot[color=thisblue, line width=1pt] table [x=date, y=Threats to Survival or Physical Security, col sep=comma] {proportions_over_time_ru.csv};

    \addplot[color=thisorange, line width=1pt] table [x=date, y=Vilification/Villainization, col sep=comma] {proportions_over_time_ru.csv};

    \addplot[color=red, line width=1pt] table [x=date, y=Explicit Dehumanization, col sep=comma] {proportions_over_time_ru.csv};

    \addplot[dashed, color=gray] coordinates {(2022-02-08,0) (2022-02-08,0.22)}
    node[pos=1, above, text=darkgray] {\scriptsize 1a}; 
    
    \addplot[dashed, color=gray] coordinates {(2022-02-21,0) (2022-02-21,0.202)}
    node[pos=1, above, text=darkgray] {\scriptsize 2a}; 
    \addplot[dashed, color=gray] coordinates {(2022-04-19,0) (2022-04-19,0.22)}
    node[pos=1, above, text=darkgray] {\scriptsize 3a}; 
    \addplot[dashed, color=gray] coordinates {(2022-06-01,0) (2022-06-01,0.22)}
    node[pos=1, above, text=darkgray] {\scriptsize 4a}; 
    \addplot[dashed, color=gray] coordinates {(2022-06-23,0) (2022-06-23,0.202)}
    node[pos=1, above, text=darkgray] {\scriptsize 5a}; 

    \addplot[dashed, color=gray] coordinates {(2022-07-08,0) (2022-07-08,0.22)}
    node[pos=1, above, text=darkgray] {\scriptsize 6a}; 

    \addplot[dashed, color=gray] coordinates {(2022-08-01,0) (2022-08-01,0.22)}
    node[pos=1, above, text=darkgray] {\scriptsize 7a}; 
    \addplot[dashed, color=gray] coordinates {(2022-09-28,0) (2022-09-28,0.22)}
    node[pos=1, above, text=darkgray] {\scriptsize 8a}; 
    \addplot[dashed, color=gray] coordinates {(2023-02-03,0) (2023-02-03,0.22)}
    node[pos=1, above, text=darkgray] {\scriptsize 9a}; 


    \end{axis}
    \end{tikzpicture}
    \vspace{-4mm}
    \\\\

    \begin{tikzpicture}
    \begin{axis}[
        xlabel=(b) Ukrainian war bloggers,
        ylabel=Proportion of Messages,
        date coordinates in=x,
        xticklabel style={rotate=90, anchor=center, xshift=-19pt}, 
        ytick={0, 0.05, 0.1, 0.15, 0.2, 0.25}, 
        yticklabels={0, 0.05, 0.1, 0.15, 0.2, 0.25},
        yticklabel style={xshift=-5pt}, 
        xtick={2021-12-01, 2022-02-01, 2022-04-01, 2022-06-01, 2022-08-01, 2022-10-01, 2022-12-01, 2023-02-01, 2023-04-01}, 
        xticklabels={Dec '21, Feb '22, Apr '22, Jun '22, Aug '22, Oct '22, Dec '22, Feb '23, Apr '23},
        xmin=2021-12-01, xmax=2023-05-01, 
        ymin=0, ymax=0.25,
        width=\linewidth,
        height=0.6\linewidth,
        axis x line*=bottom,  
        ymajorgrids=true,
        legend style={at={(0.5,-0.57)}, anchor=north,legend columns=1, font=\small}
    ]

    \addplot[color=thisgreen!90!black, line width=1pt] table [x=date, y=Threats to Culture or Identity, col sep=comma] {proportions_over_time_uk.csv};
    \addlegendentry{Threats to Culture or Identity}

    \addplot[color=thisblue, line width=1pt] table [x=date, y=Threats to Survival or Physical Security, col sep=comma] {proportions_over_time_uk.csv};
    \addlegendentry{Threats to Survival or Physical Security}

    \addplot[color=thisorange, line width=1pt] table [x=date, y=Vilification/Villainization, col sep=comma] {proportions_over_time_uk.csv};
    \addlegendentry{Vilification/Villainization}

    \addplot[color=red, line width=1pt] table [x=date, y=Explicit Dehumanization, col sep=comma] {proportions_over_time_uk.csv};
    \addlegendentry{Explicit Dehumanization}

    \addplot[dashed, color=gray] coordinates {(2022-02-08,0) (2022-02-08,0.22)}
    node[pos=1, above, text=darkgray] {\scriptsize 1b};
    
    \addplot[dashed, color=gray] coordinates {(2022-02-21,0) (2022-02-21,0.202)} 
    node[pos=1, above, text=darkgray]{\scriptsize 2b}; 


    \addplot[dashed, color=gray] coordinates {(2022-03-21,0) (2022-03-21,0.22)}
    node[pos=1, above, text=darkgray]{\scriptsize 3b}; 
    
    \addplot[dashed, color=gray] coordinates {(2022-07-03,0) (2022-07-03,0.22)}
    node[pos=1, above, text=darkgray]{\scriptsize 4b}; 
    \addplot[dashed, color=gray] coordinates {(2022-08-29,0) (2022-08-29,0.22)}
    node[pos=1, above, text=darkgray]{\scriptsize 5b}; 
    \addplot[dashed, color=gray] coordinates {(2022-09-21,0) (2022-09-21,0.22)}
    node[pos=1, above, text=darkgray]{\scriptsize 6b}; 
    \addplot[dashed, color=gray] coordinates {(2022-11-11,0) (2022-11-11,0.22)}
    node[pos=1, above, text=darkgray]{\scriptsize 7b}; 
    \addplot[dashed, color=gray] coordinates {(2022-12-29,0) (2022-12-29,0.22)}
    node[pos=1, above, text=darkgray]{\scriptsize 8b}; 
    \addplot[dashed, color=gray] coordinates {(2023-02-09,0) (2023-02-09,0.22)}
    node[pos=1, above, text=darkgray]{\scriptsize 9b}; 

    \end{axis}
    \end{tikzpicture}
    \\ 
    \end{tabular}
    \caption{Temporal trends in the proportion of messages with othering language posted by (a) \textbf{Russian war bloggers} and (b) \textbf{Ukrainian war bloggers} from December 2021 to May 2023. The four classes of othering language are: Threats to Culture or Identity, Threats to Survival or Physical Security, Vilification/Villainization, and Explicit Dehumanization.}
    \vspace{-4mm}
    \label{fig:othering_dynamics}
\end{figure}
Figure \ref{fig:othering_dynamics} shows that othering rhetoric fluctuated throughout the conflict, rising after key events. We define key events as significant events during the war that also became prominent discussion topics within their respective communities (i.e., Russian and Ukrainian war bloggers separately). Here, we utilize the events enumerated in Tables \ref{tab:events_key_ru} and \ref{tab:events_key_uk}, which were compiled using domain knowledge in conjunction with methods from previous work~\cite{gerard2024modelinginformationnarrativedetection}. 

The types of events that tend to correlate with spikes in othering rhetoric differ significantly between these communities. For Russian bloggers, the most prominent driver was US military and security aid to Ukraine, which often led to a rise in rhetoric, as they framed themselves as victims of Western aggression. In contrast, for Ukrainian bloggers, increases in othering language were primarily driven by military gains and losses on the battlefield, reflecting a more direct response to the dynamics of the war itself. These findings are consistent with previous research showing that Russian bloggers tend to react more to international actions, while Ukrainian bloggers are more focused on military developments \cite{gerard2024modelinginformationnarrativedetection}. In future work, we plan to automate event detection using change point analysis and further explore the causal relationship between these events and the prevalence of othering language.

\vspace{-1mm}
\subsection{The Moral Language of Othering}


We explore the interaction between othering and moral language on Telegram and Gab. For both platforms, we label moral values expressed in text using a model trained to recognize moral language~\cite{trager2022moral}. The model identifies the moral foundations of human intuitive ethics, such as valuing of purity, respect for authority, equality (fairness), group loyalty, and care or protection of the more vulnerable~\cite{graham2013moral}. 



\subsubsection{Moralized Othering on Telegram}
We begin by analyzing messages posted by war bloggers on Telegram, using a chi-squared test to explore the relationship between moral language in messages containing othering language versus those without it. The chi-squared test helps determine whether the association between moral language and othering is statistically significant, rather than occurring by chance. The results indicate a strong, statistically significant connection ($p < 0.001$), demonstrating that moral language is more likely to co-occur with othering language than with non-othering language. This suggests that moral framing is frequently used to justify or intensify othering language, supporting the Moralized Threat Hypothesis~\cite{hoover2021investigating}.

\begin{figure*}[t] 
    \centering
    \begin{subfigure}[t]{0.49\textwidth} 
        \centering
        \includegraphics[width=\textwidth]{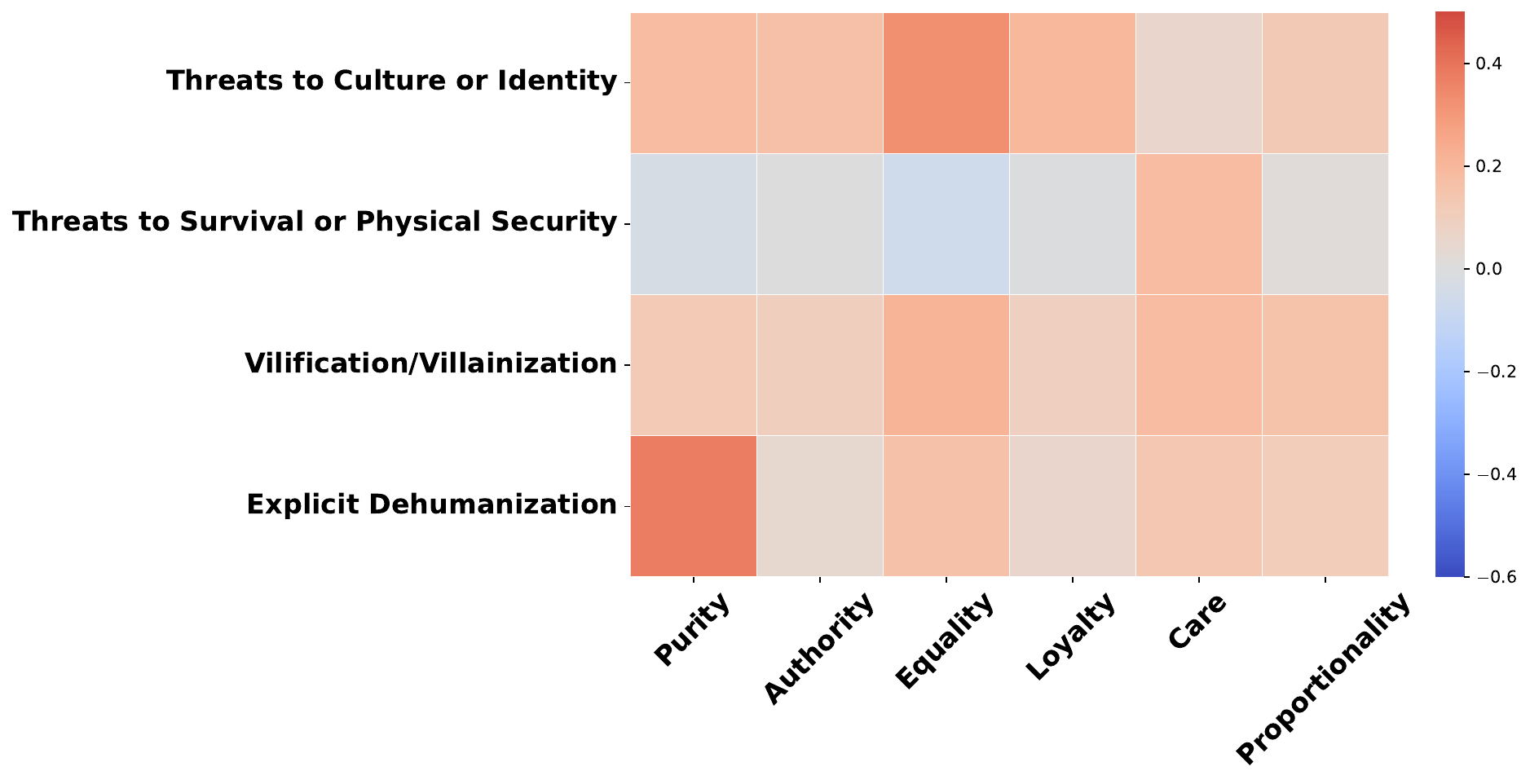} 
        \caption{Log-odds ratios for morality devices in \textbf{Russian war bloggers}' messages.}
        \label{fig:log_odds_ru}
    \end{subfigure}
    \hfill
    \begin{subfigure}[t]{0.49   \textwidth} 
        \centering
        \includegraphics[width=\textwidth]{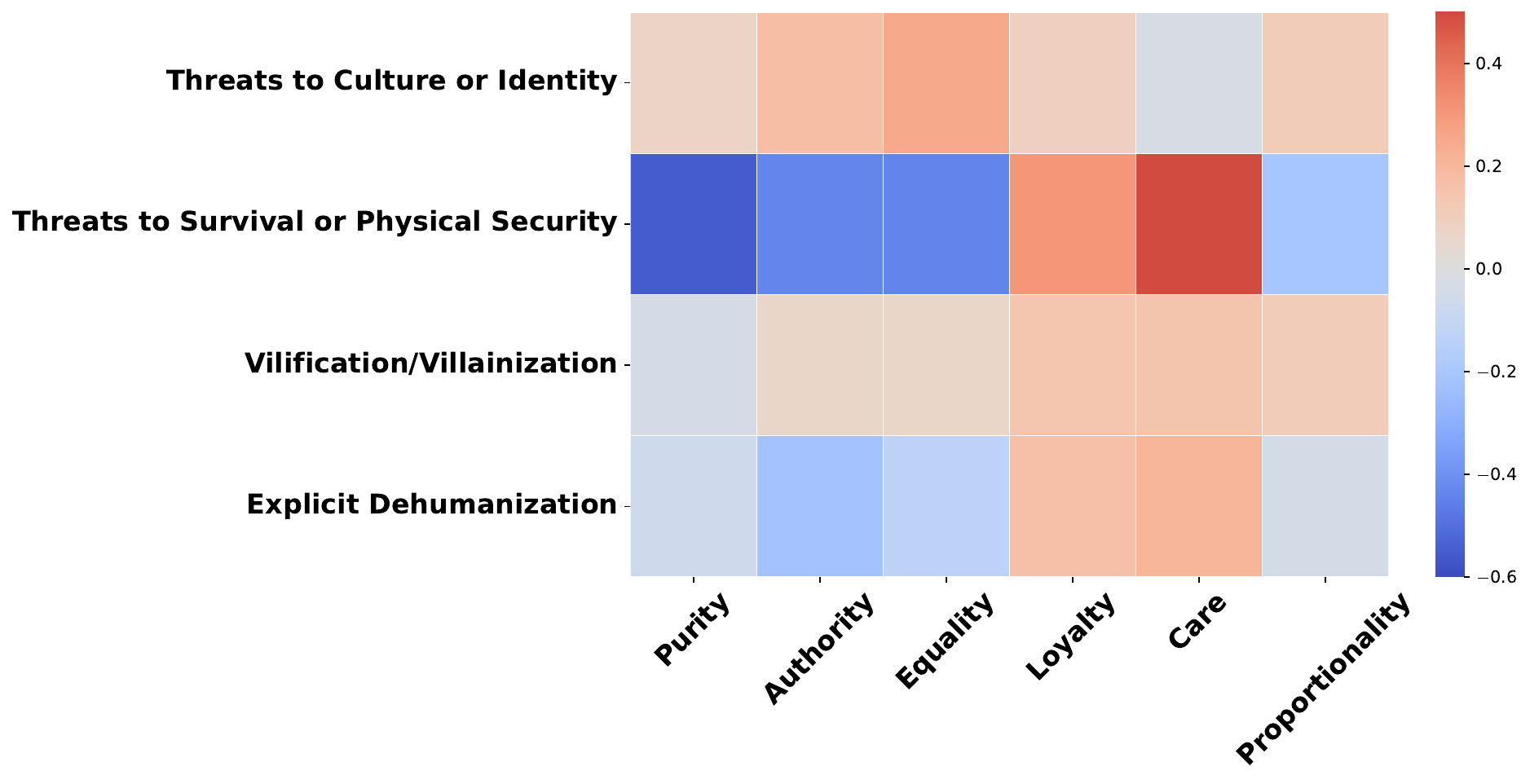} 
        \caption{Log-odds ratios for morality devices in \textbf{Ukrainian war bloggers}' messages.}
        \label{fig:log_odds_uk}
    \end{subfigure}
    \caption{Comparison of log-odds ratios for morality devices across othering categories in Russian and Ukrainian war bloggers' messages. The color intensity reflects the strength of the association, with warmer colors (red) indicating higher positive log-odds ratios and cooler colors (blue) representing negative or lower values.}
    \label{fig:log_odds_comparison_heatmap}
\end{figure*}

Next, we analyzed the use of moral language by Russian and Ukrainian war bloggers across different othering categories to identify differences in their moral framing strategies. Figure \ref{fig:log_odds_otherism_categories} highlights the significant variations in the moral values expressed by each side when using othering language, based on log-odds ratios. We also examined the interaction between specific moral language categories and individual othering classes. As shown in Figure \ref{fig:log_odds_comparison_heatmap}, the two groups also differ significantly in their use of moral language within othering rhetoric, with these differences confirmed by two-proportion z-tests ($p < 0.001$). 

Explicit forms of othering, such as \textit{Explicit Dehumanization}, are the least associated with moral language in both groups, likely due to their overtly aggressive nature, which doesn't align well with broader moral frameworks \cite{saha2023rise}. However, for Russian war bloggers, the strongest association between morality and othering is found in the purity moral frame within \textit{Explicit Dehumanization}. This suggests that while Russians use moral language less frequently with dehumanization, when they do, it is often tied to purity, reflecting popular narratives portraying Ukrainians (and the West) as `puppets' or `zombies' corrupting Russian values. Meanwhile, for Ukrainian war bloggers, the most morally charged category is \textit{Threats to Survival or Physical Safety}, most closely associated with care, which aligns with the context of Russia's invasion.


Overall, both groups display similar trends, but the use of moral language reveals strategic differences. Russian bloggers emphasize purity and cultural threats, reinforcing existential threat and victimization~\cite{geissler2023russian}, while Ukrainian bloggers focus on care in response to physical threats. Our analysis shows that moral language and othering are deeply intertwined, with Russian bloggers heavily relying on moralized language to justify intergroup prejudice. This suggests moralized othering is an effective propaganda tool, though further research is needed to understand its role in polarized environments.

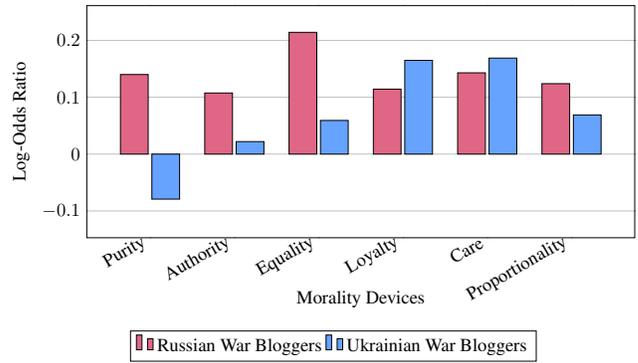
\begin{figure}[h]
    \centering
    \resizebox{\linewidth}{!}{%
    \begin{tikzpicture}
        \begin{axis}[
            ybar,
            width=12cm,
            height=6cm,
            symbolic x coords={Purity, Authority, Equality, Loyalty, Care, Proportionality},
            xtick=data,
            xlabel={Morality Devices},
            ylabel={Log-Odds Ratio},
            ymin=-0.1,
            bar width=15pt,
            enlargelimits=0.15,
            xlabel style={yshift=15pt},  
            x tick label style={rotate=30, anchor=east},
            legend style={at={(0.45,-0.4)}, anchor=north, legend columns=-1},
            legend cell align={left},
            legend entries={Russian War Bloggers, Ukrainian War Bloggers},
            every axis/.append style={
                ytick scale label code/.code={},
            },
            ymajorgrids=true,
            tick style={
                major tick length=0pt, 
            },  
        ]
            \addplot[fill=red!60!white] coordinates {
                (Purity, 0.1400)
                (Authority, 0.1075)
                (Equality, 0.2142)
                (Loyalty, 0.1143)
                (Care, 0.1430)
                (Proportionality, 0.1239)
            };
            
            \addplot[fill=blue!60!white] coordinates {
                (Purity, -0.0792)
                (Authority, 0.0220)
                (Equality, 0.0592)
                (Loyalty, 0.1649)
                (Care, 0.1689)
                (Proportionality, 0.0689)
            };
            
        \end{axis}
    \end{tikzpicture}
    }
    \caption{Log-odds ratios of morality devices for Russian and Ukrainian war bloggers, comparing the presence of moral language in messages with othering language versus those without. Ukrainian war bloggers are represented in light blue, and Russian war bloggers in light red.}
    \label{fig:log_odds_morality_devices}
\end{figure}

\begin{figure}[h]
\vspace{-2mm}
    \hspace{-3mm}
    \resizebox{0.45\textwidth}{!}{%
    \begin{tikzpicture}
        \begin{axis}[
            ybar,
            width=10cm,
            height=6cm,
            symbolic x coords={Culture or Identity, Physical Survival or Security, Villainization, Dehumanization},
            xtick=data,
            xlabel={Othering Categories},
            ylabel={Log-Odds Ratio},
            ymin=0,
            bar width=15pt,
            enlargelimits=0.15,
            xlabel style={yshift=10pt},  
            x tick label style={rotate=30, anchor=east},
            legend style={at={(0.45,-0.62)}, anchor=north, legend columns=-1},
            legend cell align={left},
            legend entries={Russian War Bloggers, Ukrainian War Bloggers},
            every axis/.append style={
                ytick scale label code/.code={},
            },
            ymajorgrids=true,
            tick style={
                major tick length=0pt, 
            },  
        ]
            \addplot[fill=red!60!white] coordinates {
                (Culture or Identity, 0.1754)
                (Physical Survival or Security, 0.1830)
                (Villainization, 0.2295)
                (Dehumanization, 0.096)
            };
            
            \addplot[fill=blue!60!white] coordinates {
                (Culture or Identity, 0.0887)
                (Physical Survival or Security, 0.2046)
                (Villainization, 0.1576)
                (Dehumanization, 0.0764)
            };
            
        \end{axis}
    \end{tikzpicture}
    }
    \caption{Log-odds ratios of morality use for Russian and Ukrainian war bloggers, comparing its use in messages with specific othering language use versus those without. Ukrainian war bloggers are represented in light blue, and Russian war bloggers are represented in light red.}
    \label{fig:log_odds_otherism_categories}
\end{figure}
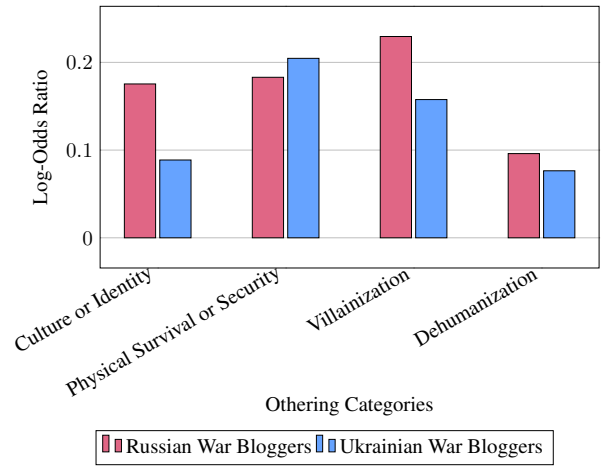

\begin{figure}[h]
    \hspace{-3mm}
    \resizebox{0.45\textwidth}{!}{%
    \begin{tikzpicture}
        \begin{axis}[
            ybar,
            width=10cm,
            height=6cm,
            symbolic x coords={Purity, Authority, Equality, Loyalty, Care, Proportionality},
            xtick=data,
            xlabel={Morality Devices},
            ylabel={Log-Odds Ratio},
            ymin=0.0,
            bar width=15pt,
            enlargelimits=0.15,
            xlabel style={yshift=10pt},  
            x tick label style={rotate=30, anchor=east},
            every axis/.append style={
                ytick scale label code/.code={},
            },
            ymajorgrids=true,
            tick style={
                major tick length=0pt, 
            },  
        ]
            \addplot[fill=mypurple] coordinates {
                (Purity, 0.5268)
                (Authority, 0.3182)
                (Equality, 0.5359)
                (Loyalty, 0.2096)
                (Care, 0.4926)
                (Proportionality, 0.3406)
            };
            
        \end{axis}
    \end{tikzpicture}
    }
    \caption{Log-odds ratios of morality devices for Gab, comparing the presence of moral language in messages with othering language versus those without.}
    \vspace{-3mm}
    \label{fig:log_odds_morality_devices_gab}
\end{figure}
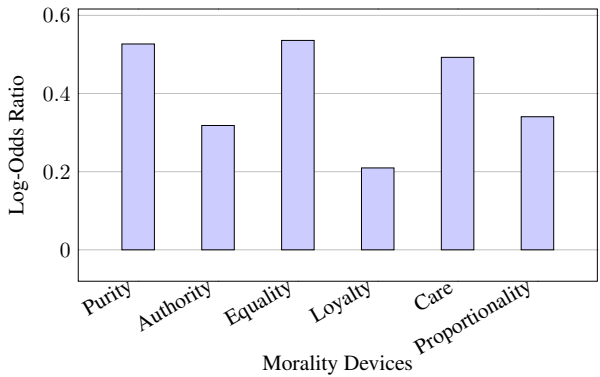

\subsubsection{Moralized Othering on Gab}

Gab provides an additional lens for examining the interplay between moral language and othering in a polarized environment with lower levels of immediate conflict. Although not directly involved in intergroup violence, the platform's rhetoric is steeped in othering narratives within an American context. This makes Gab an ideal setting to further test our hypothesis in a context similarly shaped by divisive and exclusionary discourse, but without direct conflict.

\begin{figure} 

\hspace{-6mm} 
\includegraphics[width=0.52\textwidth]{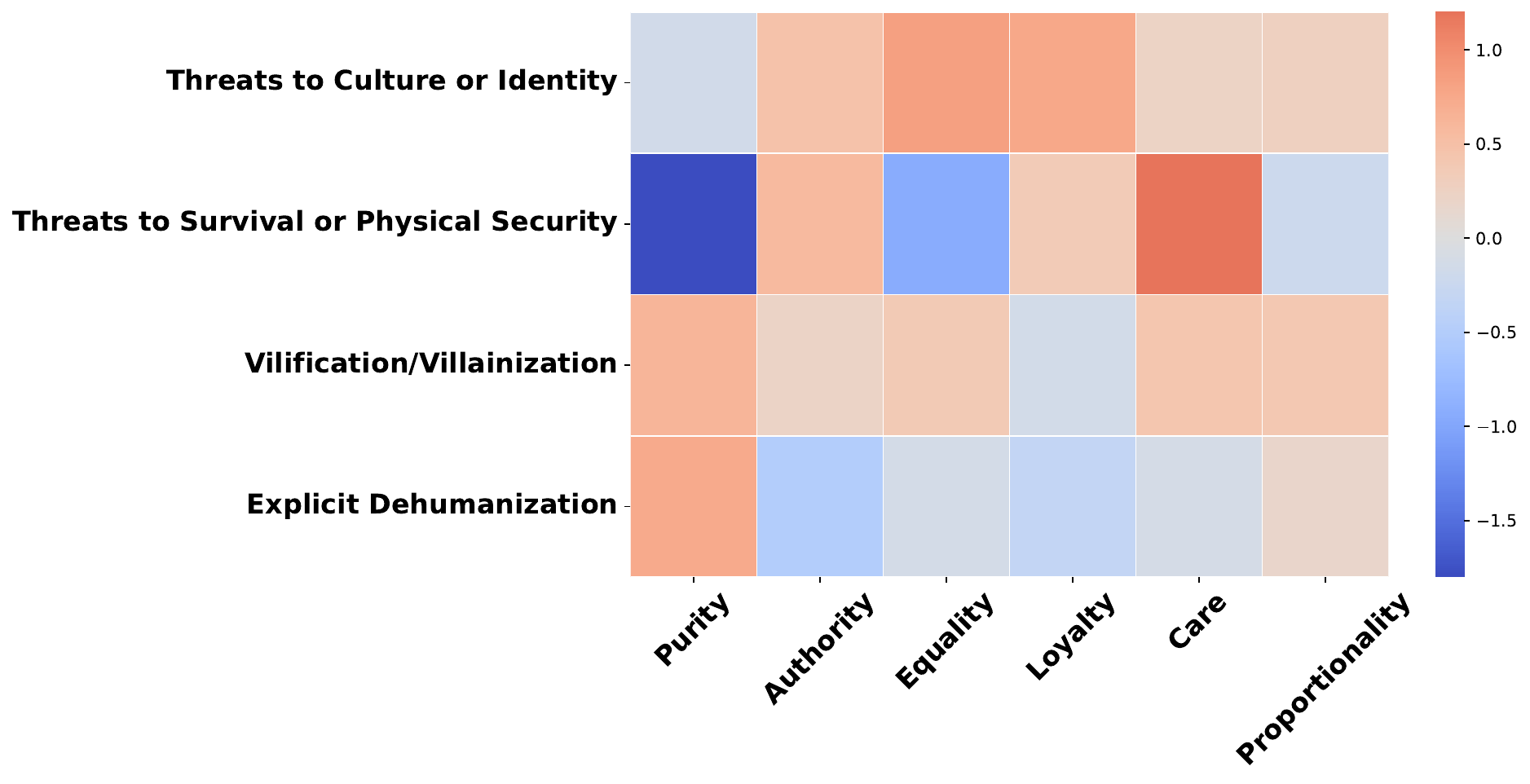} 
\caption{Heatmap displays the log-odds ratios for the use of various morality devices (Purity, Authority, Equality, Loyalty, Care, and Proportionality) across different othering categories (Threats to Culture or Identity, Threats to Survival or Physical Security, Vilification/Villainization, and Explicit Dehumanization) in the \textbf{Gab users'} messages. The color intensity reflects the strength of the association, with warmer colors (red) indicating higher positive log-odds ratios and cooler colors (blue) representing negative or lower values.}
\label{fig:log_odds_gab_heatmap}
\end{figure}

As in the previous case study, we first applied the chi-squared test to confirm a significant difference in the use of moral language between messages with and without othering language ($p < 0.001$). We then evaluated the log-odds ratios of moral language in othering messages and the interaction between specific moral categories and othering types, as shown in Figures\ref{fig:log_odds_morality_devices_gab} and \ref{fig:log_odds_gab_heatmap}. The most morally loaded othering category is \textit{Threats to Survival and Physical Security}, primarily associated with care, similar to Ukrainians. \textit{Threats to Culture or Identity} is linked to equality and loyalty, similar to Russians. \textit{Explicit Dehumanization} is tied to purity but not to other moral values, unlike in Russians, where it had (weak) links to all moral values.

These results not only support the hypothesis that moral language and othering are closely intertwined but also reaffirm the findings from the previous case study, highlighting how moral language shapes and sustains othering narratives. They also offer a new lens on how different moral devices are strategically used to appeal to different audiences, underscoring the need for further research into how these narratives are constructed and spread in polarized environments.

\definecolor{mygreen}{RGB}{ 158, 219, 169 }
\definecolor{mypurple}{RGB}{ 187, 158, 219 }


\subsection{Othering and Attention}

Next, we explore the relationship between online attention and the use of othering language in messages by Russian and Ukrainian war bloggers. 

\subsubsection{Network Centrality}
We measure channel centrality in the reference network of war blogger channels (see Methods) using degree centrality and eigenvector centrality. (For simplicity, we use an undirected version of this network.) Degree centrality reflects the influence distribution and communicative ability of nodes in the network
and eigenvector centrality captures the positional importance of network nodes. We then calculate the Spearman correlation between channel centrality and its use of othering language (as a proportion of its messages). 

The results, shown in Table \ref{tab:network_centrality}, reveal statistically significant correlations for both degree centrality and eigenvector centrality and the use of othering language by both groups of war bloggers, with a stronger correlation among Russian war bloggers.
This suggests that war bloggers who use othering language occupy more influential positions within the Telegram network, though the reasons remain unclear. It could be that existing opinion leaders are more inclined to use inflammatory othering language, or that such language attracts more attention, rewarding those who use it. This finding does not disentangle these possibilities.


\subsubsection{Message Views}
\begin{figure}[h]
\vspace{-4mm}
\resizebox{0.45\textwidth}{!}{%
    \begin{tikzpicture}
        \begin{axis}[
            ybar,
            width=8cm,
            height=6cm,
            symbolic x coords={Russian war bloggers, Ukrainian war bloggers},
            xtick=data,
            xlabel={Groups},
            ylabel={Mean Views (Normalized by Channel)},
            ymin=-0.05,
            ymax=0.11,
            ytick={-0.05, 0, 0.05, 0.1}, 
            yticklabels={-0.05, 0, 0.05, 0.1}, 
            bar width=20pt,
            axis line style={line width=.3pt}, 
            x tick label style={yshift=-15pt, anchor=center}, 
            tick style={
                major tick length=0pt, 
            },
            legend style={at={(0.5,-0.3)}, anchor=north, legend columns=-1},
            legend cell align={left},
            legend entries={With othering, Without othering},
            ymajorgrids=true,
            enlarge x limits={abs=1.00cm}, 
        ]
            \addplot[fill=mygreen!70!white, error bars/.cd, y dir=both, y explicit]
            coordinates {
                (Russian war bloggers, 0.094) +- (0, 0.0092)
                
                (Ukrainian war bloggers, 0.013) +- (0, 0.0024)
            };
            
            \addplot[fill=mypurple!80!white, error bars/.cd, y dir=both, y explicit]
            coordinates {
                (Russian war bloggers, -0.0197) +- (0, 0.0028)
                
                (Ukrainian war bloggers, -0.0045) +- (0, 0.0013)
            };
        \end{axis}
    \end{tikzpicture}
    }
    \caption{Comparison of mean views with and without othering (z-score channel-normalized). The bars represent the mean views for Russian and Ukrainian war bloggers, with and without othering, and the error bars indicate the standard error.}
    \label{fig:mean_views_otherism}
    \vspace{-4mm}
\end{figure}
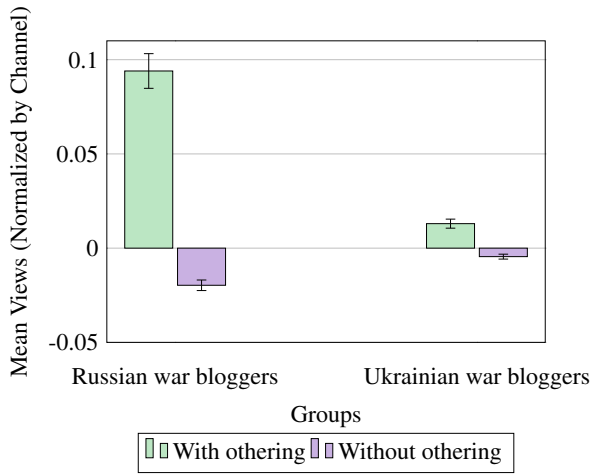

To further investigate the link between othering and attention, we focus on the messages themselves. We normalize each message's number of views by the channel's typical viewership using Z-Score normalization. As shown in Figure \ref{fig:mean_views_otherism}, messages containing othering language consistently garner more views than those without, a relationship confirmed by the Mann-Whitney U Test ($p < 0.001$), which demonstrates a statistically significant difference between the number of views of messages with and without othering language.
These results show that othering messages attract more attention than regular messages.


\subsubsection{Times of Crisis}
We define times of crisis as the time period immediately following significant events (one week) during the war that are popular discussion topics amongst the community (in this case Russian and Ukrainian war bloggers separately). Here, we utilize the events enumerated in Tables \ref{tab:events_key_ru} and \ref{tab:events_key_uk}. Overall, we find that othering receives more attention immediately following a crisis. 

\begin{figure}[h]
\resizebox{0.5\textwidth}{!}{%

    \begin{tikzpicture}
        \begin{axis}[
            ybar,
            width=8cm,
            height=6cm,
            symbolic x coords={Russian war bloggers, Ukrainian war bloggers},
            xtick=data,
            xlabel={Groups},
            ylabel={Mean Views (Normalized by Channel)},
            ymin=-0.05,
            ymax=0.1,
            ytick={-0.05, 0, 0.05, 0.1}, 
            yticklabels={-0.05, 0, 0.05, 0.1}, 
            bar width=20pt,
            axis line style={line width=.3pt}, 
            x tick label style={yshift=-15pt, anchor=center}, 
            tick style={
                major tick length=0pt, 
            },
            legend style={at={(0.5,-0.29)}, anchor=north, legend columns=-1},
            legend cell align={left},
            legend entries={With othering, Without othering},
            ymajorgrids=true,
            enlarge x limits={abs=1.00cm}, 
        ]
            \addplot[fill=mygreen!70!white, error bars/.cd, y dir=both, y explicit]
            coordinates {
                (Russian war bloggers, 0.067170) +- (0, 0.028656)
                (Ukrainian war bloggers, 0.035249) +- (0, 0.007124)
            };
            
            \addplot[fill=mypurple!80!white, error bars/.cd, y dir=both, y explicit]
            coordinates {
                (Russian war bloggers, -0.012976) +- (0, 0.010005)
                (Ukrainian war bloggers, -0.013410) +- (0, 0.003974)
            };
        \end{axis}
    \end{tikzpicture}
    }
    \caption{Comparison of mean views with and without othering (z-score channel-normalized) following crises. The bars represent the mean views for Russian and Ukrainian war bloggers, with and without othering, and the error bars indicate the standard error.}
    \label{fig:mean_views_otherism_inflamed_period}

\end{figure}
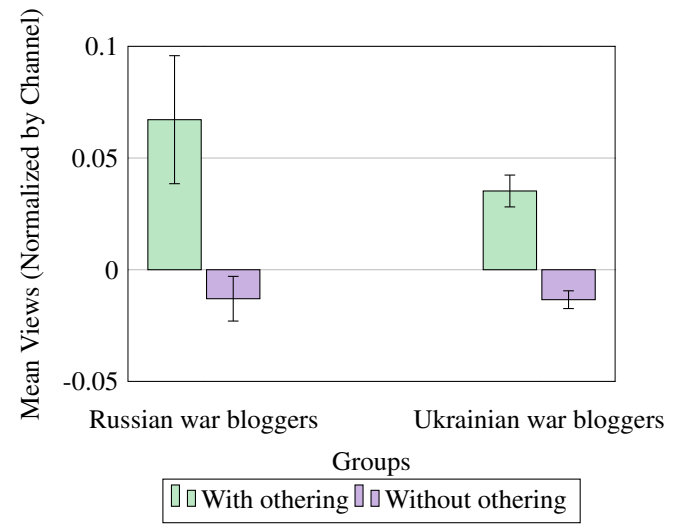

As shown in Figure \ref{tab:network_centrality_following_events}, the correlation between degree centrality and the proportion of messages containing othering increases markedly following events. For Russian war bloggers, there is also a significant rise in the correlation between eigenvector centrality and othering, while for Ukrainian war bloggers, there is a slight decrease. Additionally, Figure \ref{fig:mean_views_otherism_inflamed_period} illustrates a shift in viewership dynamics. For Ukrainian war bloggers, the gap in viewership between messages with and without othering widens dramatically during periods of heightened conflict, a statistically significant change confirmed by the Mann-Whitney U Test. In contrast, the viewership differential for Russian war bloggers remains relatively stable, even during crises. These findings suggest that while othering generally increases attention, its impact becomes particularly pronounced in times of crisis.

This analysis highlights that othering not only correlates with increased attention in online discourse but may also be structurally rewarded, especially during crises. Both network centrality and viewership metrics suggest that users engaging in othering are more likely to occupy influential positions within their networks and enjoy greater visibility. The sharp increase in attention during periods of heightened conflict, particularly among Ukrainian war bloggers, emphasizes the role of othering in driving engagement. These findings strongly suggest that othering functions as a powerful tool for capturing attention and influence in polarized environments, with its effects significantly amplified in times of crisis.

\begin{table}[h]
\vspace{-4mm}
\centering
\small
\resizebox{0.32\textwidth}{!}{%
\begin{tabular}{lcc}
\textbf{} & \multicolumn{2}{c}{\textbf{\makebox[0pt][c]{Centrality Metric}}} \\ 
\cline{2-3} \addlinespace[1mm]
Community & Degree & Eigenvector \\
\hline
\addlinespace[1mm]
Russian & 0.254 & 0.333 \\ 
Ukrainian & 0.128 & 0.147 \\
\hline
\end{tabular}
}
\caption{Centrality and othering messages. Spearman correlation between a channel's proportion of messages with othering language and its degree and eigenvector centralities. All correlations are significant at the $p < 0.01$ level.}
\label{tab:network_centrality}
\vspace{-3mm}
\end{table}

\vspace{-4mm}
\section{Discussion and Conclusion}
Our work presents a comprehensive taxonomy of othering language, grounded in sociological theory and contextualized alongside established concepts like hate speech and fear speech. This framework offers a structured approach to identifying and understanding othering language, clarifying its role in online discourse. We also introduce methods for efficiently training LLMs to detect othering language and transfer this knowledge to new domains, aiming to help advance future efforts in understanding and moderating it.

Our analysis highlights the dynamic nature of othering language and its responsiveness to real-world events. We establish a clear link between othering and moral language, reinforcing the Moralized Threat Hypothesis and showing how moral framing can amplify harmful narratives. Additionally, we find a strong connection between othering language and social attention, with such messages receiving higher engagement, especially during times of crisis when othering intensifies and attracts disproportionately more attention.

\vspace{-2mm}
\subsection{Ethical Considerations}
\vspace{-1mm}
Our study used publicly accessible data from influential figures discussing the Russia-Ukraine war on Telegram, following the FAIR data principles. While we aim to advance computational analysis, we recognize the risks of misuse, such as evading classifiers or generating harmful language; applying our classifiers poses its own challenges: over-detection may suppress legitimate speech, while under-detection can spread harmful rhetoric. To ensure transparency and responsible use, we have made both the data and code publicly available, stressing the ethical application of these resources.
\vspace{-2mm}
\subsection{Limitations}
\vspace{-1mm}
Our analysis is centered on Russian and Ukrainian war bloggers operating in a highly charged war environment, which may impact the neutrality of the data. Additionally, while the study spans the first year and a half of the conflict, the war is ongoing, and future developments may alter the narrative dynamics. The centrality metrics we used are based on networks generated over the entire timeframe, which limits their dynamism compared to platforms like Twitter, where more edges are created. Finally, while our model allows for editing the system prompt in the RDA process, many other open-source models, such as Mistral, do not offer this flexibility, which may limit adaptability in similar studies.

In addition, biases could inadvertently find their way into our results. One source of bias is moral annotation process, which was trained on out-of-domain data. However, when doing comparative analysis, biases should largely cancel.
\vspace{-1mm}
\subsection{Summary and Future Work}
This study analyzes how othering language is used by Russian and Ukrainian war bloggers on Telegram, highlighting its dynamic evolution in response to external events and its interaction with moral language and social attention. We develop a taxonomy for othering language and introduce methods for training LLMs to detect and transfer this knowledge to new domains. Our findings show that othering intensifies during crises, attracting more social attention and rewards, complicating the online discourse.

Future work should extend these methods to other conflicts and domains to test their generalizability, providing insights into how othering and social attention interact in polarized environments. Additionally, future research should explore how proximity to the outgroup shapes othering language. Specifically, core members may rely more on ideological and coded language to reinforce identity, while boundary members tend to adopt defensive and confrontational rhetoric, justifying aggression and emphasizing immediate threats. Understanding these dynamics could reveal further nuances in the strategic use of moralized othering.

\bibliography{LaTeX/aaai24}

\section{Appendix}

\begin{table}[h]
\resizebox{0.5\textwidth}{!}{%

\centering
\begin{tabular}{lc}
Category & Instance Counts 
\\\addlinespace[1mm]
\hline
Threats to Culture or Identity & 122 \\ 
Threats to Survival or Physical Security & 62 \\ 
Vilification/Villainization & 164 \\ 
Explicit Dehumanization & 77 \\ 
None & 78 \\ \hline
\addlinespace[1mm]
\textbf{Total Data Points} & \textbf{316} \\ \hline
\end{tabular}
}
\caption{Summary of human annotations for Russian war bloggers data.}
\label{tab:gold_set_summary}
\end{table}

\begin{table}[h]

\resizebox{0.5\textwidth}{!}{%

\begin{tabular}{lcc}
\\\addlinespace[1mm]

Category & Krippendorff's & Fleiss'
\\
\addlinespace[1mm]
\hline
Threats to Culture or Identity & 0.72 & 0.72\\ 
Threats to Survival or Physical Security & 0.62 & 0.62 \\ 
Vilification/Villainization & 0.70 & 0.73\\ 
Explicit Dehumanization & 0.68 & 0.65\\ 
None & 0.70 & 0.73\\ \hline
\end{tabular}
}
\caption{Inter-Annotator Agreement: Krippendorff's Alpha and Fleiss' Kappa for Russian war bloggers data.}
\label{tab:krippendorff_alpha}
\end{table}

\begin{table}[h]

\centering
\resizebox{0.5\textwidth}{!}{%
\small
\begin{tabular}{lcccc}
\noalign{\smallskip} 
Category & Cohen's & Accuracy & F1\\\addlinespace[1mm]
 \hline
Threats to Culture or Identity & 0.83 & 0.92 & 0.92\\ 
Threats to Survival/Security & 0.75 &  0.82 & 0.80\\ 
Vilification/Villainization & 0.80 & 0.90 & 0.90\\ 
Explicit Dehumanization & 0.85 & 0.97 & 0.97 \\ 
None & 0.80 & 0.92 & 0.92\\ \hline
\end{tabular}
}
\caption{Inter-Annotator Agreement and Model Performance: Cohen’s Kappa (Agreement), Accuracy, and F1 Score between majority vote and HQ-LLM (ChatGPT-4o) on Russian war bloggers data.}
\label{tab:accuracy_f1_ru_gpt}
\end{table}

\begin{table}[h]
\centering
\small
\begin{tabular}{lc}
Category & Instance Counts \\\addlinespace[1mm]

\hline
Threats to Culture or Identity & 45 \\ 
Threats to Survival or Physical Security & 41 \\ 
Vilification/Villainization & 52 \\ 
Explicit Dehumanization & 32 \\ 
None & 87 \\ \hline
\addlinespace[1mm]

\textbf{Total Data Points} & \textbf{212} \\ \hline
\end{tabular}
\caption{Human-annotated gold set summary for Ukrainan war bloggers data.}
\label{tab:ukrainian_gold_set_summary}
\end{table}

\begin{table}[h]
\centering
\resizebox{0.5\textwidth}{!}{%
\small
\begin{tabular}{lcc}
Category & Cohen's \\\addlinespace[1mm]
\hline
Threats to Culture or Identity & 0.79 \\ 
Threats to Survival or Physical Security & 0.77 \\ 
Vilification/Villainization & 0.79 \\ 
Explicit Dehumanization & 0.74 \\ 
None & 0.73 \\ \hline
\end{tabular}
}
\caption{Inter-Annotator Agreement: Cohen's Kappa for Ukrainian war bloggers data.}
\label{tab:ukrainian_gold_set_summary}
\end{table}

\begin{table}[h]
\centering
\resizebox{0.5\textwidth}{!}{%

\begin{tabular}{lccc}

Category & Cohen's & Accuracy & F1 Score \\\addlinespace[1mm]

\hline

Threats to Culture or Identity & 0.80 & 0.90 & 0.91 \\ 
Threats to Survival or Physical Security & 0.76 & 0.81 & 0.82 \\ 
Vilification/Villainization & 0.78 & 0.89 & 0.88 \\ 
Explicit Dehumanization & 0.81 & 0.96 & 0.96 \\ 
None & 0.83 & 0.93 & 0.92 \\ \hline
\end{tabular}
}
\caption{Inter-Annotator Agreement and Model Performance: Cohen’s Kappa (Agreement), Accuracy, and F1 Score between majority vote and HQ-LLM (ChatGPT-4o) on Ukrainian war bloggers data.}
\label{tab:accuracy_f1_uk_gpt}
\end{table}

\begin{table}[h]
\centering
\small
\begin{tabular}{lc}

Category & Instance Counts \\\addlinespace[1mm]

\hline
Threats to Culture or Identity & 120 \\ 
Threats to Survival or Physical Security & 74 \\ 
Vilification/Villainization & 136 \\ 
Explicit Dehumanization & 35 \\ 
None & 114 \\ \hline
\addlinespace[1mm]
\textbf{Total Data Points} & \textbf{329} \\ \hline
\end{tabular}
\caption{Summary of human annotations for Gab data.}
\label{tab:gab_gold_set_summary}
\end{table}

\begin{table}[h]
\centering
\small
\resizebox{0.5\textwidth}{!}{%
\begin{tabular}{lcc}
Category & Cohen's \\\addlinespace[1mm]

\hline

Threats to Culture or Identity & 0.87 \\ 
Threats to Survival or Physical Security & 0.88 \\ 
Vilification/Villainization & 0.88 \\ 
Explicit Dehumanization & 0.92 \\ 
None & 0.91 \\ \hline
\end{tabular}
}
\caption{Inter-Annotator Agreement: Cohen's Kappa for Gab data.}
\label{tab:gab_gold_set_summary}
\end{table}

\begin{table}[h]
\vspace{-2mm}
\centering
\small
\resizebox{0.5\textwidth}{!}{%
\begin{tabular}{lccc}
\textbf{} & \multicolumn{3}{c}{\textbf{Dataset F1 Score}} \\
\addlinespace[1mm]
\cline{2-4} \addlinespace[1mm]

Prompting Type & Russian & Ukrainian & Gab \\
\hline
\addlinespace[1mm]
No Additional Prompt & 0.74 & 0.66 & 0.53 \\ 
In-Context Learning & 0.72 & 0.63 & 0.63 \\
System Prompt & \textbf{0.78} & 0.75 & 0.76 \\ 
RDA & \textbf{0.78} & \textbf{0.76} & \textbf{0.77} \\ 
\hline
\end{tabular}
}
\caption{F1-Score comparison across different prompting types and test sets (Russian Dataset, Ukrainian Dataset, Gab Dataset).}
\label{tab:f1_score_comparison}
\end{table}

\pgfplotstableread[col sep=comma]{filtered_metrics_by_class.csv}\datatable
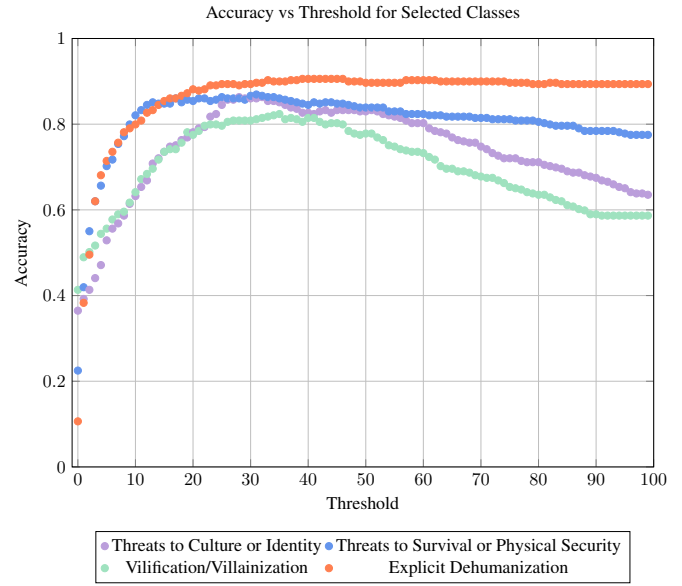
\begin{figure}
    \centering
    \resizebox{0.5\textwidth}{!}{%
    \begin{tikzpicture}
        \begin{axis}[
            title={Accuracy vs Threshold for Selected Classes},
            xlabel={Threshold},
            ylabel={Accuracy},
            grid=major,
            xmin=-1, xmax=100,  
            ymin=0, ymax=1,  
            legend style={
                at={(0.5,-0.15)},
                anchor=north,
                legend columns=2,
                align=left        
            },
            width=13cm,
            height=10cm,
            scatter/classes={
                {Threats to Culture or Identity}={mypurple, mark=*},
                {Threats to Survival or Physical Security}={thisblue, mark=*},
                {Vilification/Villainization}={thisgreen, mark=*},
                {Explicit Dehumanization}={thisorange, mark=*}
            }
        ]
        
        \addplot[scatter, only marks, 
            scatter src=explicit symbolic] 
            table[x=Threshold, y=Accuracy, meta=Class] 
            {\datatable};

        \legend{Threats to Culture or Identity, 
                Threats to Survival or Physical Security, 
                Vilification/Villainization, 
                Explicit Dehumanization}
        \end{axis}
    \end{tikzpicture}
    }
\caption{Scatter plot of accuracy vs. threshold for selected classes. This plot illustrates how adjusting confidence thresholds for classification logits affects accuracy across different classes in a new domain (Gab corpus), using a model initially trained on Russian war bloggers.}
\label{fig:logit_disambiguation}
\vspace{-2mm}
\end{figure}

\definecolor{thisgreen}{RGB}{123, 203, 162}
\begin{table}[h]
\centering
\resizebox{0.42\textwidth}{!}{%
\begin{tabular}{lcc}
\textbf{} & \multicolumn{2}{c}{\textbf{\makebox[0pt][c]{Centrality Metric}}} \\ 
\cline{2-3} \addlinespace[1mm]

Community & Degree & Eigenvector \\
\hline
\addlinespace[1mm]
Russian & 0.290 (+13.2\%) & 0.385 (+14.5\%)\\ 
Ukrainian & 0.177 (+32.1\%) & 0.136 (-7.8\%)\\
\hline
\end{tabular}
}
\caption{Centrality and othering messages following key events. Spearman correlation between a channel's proportion of messages with othering language and its degree and eigenvector centralities. All correlations are significant at the $p < 0.01$ level.}
\label{tab:network_centrality_following_events}
\end{table}

\begin{table*}[t]
\centering
\small
\begin{tabular}{m{0.1\textwidth}p{0.75\textwidth}m{0.1\textwidth}}
\textbf{Date} & \textbf{Event} & \textbf{Key} \\ 
\hline
\addlinespace[1mm]
2022-02-08 & Putin claims allowing Ukraine to join NATO would increase the prospects of a Russia-NATO conflict that could turn nuclear. & 1a \\ 
\hline
\addlinespace[1mm]2022-02-21 & Putin cites Nazism in Ukraine in speech legitimizing upcoming invasion.
 & 2a \\ 
\hline
\addlinespace[1mm]
2022-02-24 & Russia invades Ukraine.
 & - \\ 
\hline
\addlinespace[1mm]
2022-04-19 & Russia officially pivots to `next phase' of war. Russia shifted its troops from the Kyiv offensive to Ukraine's eastern Donbas region, and the amassed forces launched a broad attack there on April 18. Ukraine called it a ``new phase of the war.''
 & 3a \\ 
\hline
\addlinespace[1mm] 
2022-06-01 & The Biden administration authorizes an 11th presidential drawdown of security assistance to Ukraine valued at up to \$700 million.
 & 4a \\ 
\hline
\addlinespace[1mm] 
2022-06-23 & The Biden administration authorizes a 13th presidential drawdown of security assistance to Ukraine valued at up to \$450 million.
 & 5a \\ 
\hline
\addlinespace[1mm] 
2022-07-08 & The Biden administration announces \$400 million in additional security assistance for Ukraine.
 & 6a \\ 
\hline
\addlinespace[1mm] 
2022-08-01 & The Biden administration announces \$550 million in additional security assistance for Ukraine.
 & 7a \\ 
\hline
\addlinespace[1mm] 
2022-09-28 & United States Department of Defense announces approximately \$1.1 billion in additional security assistance for Ukraine.
 & 8a \\ 
\hline
\addlinespace[1mm] 
2023-02-03 & United States Department of Defense announces a significant new package of security assistance for Ukraine, including the authorization of a presidential drawdown of security assistance valued at up to \$425 million, as well as \$1.75 billion in Ukraine Security Assistance Initiative (USAI) funds.
 & 9a \\ 
\hline
\end{tabular}
\caption{Key events in the war discussed by Russian war bloggers. The ``Key'' column corresponds to the labeled vertical lines in Figure \ref{fig:othering_dynamics}. Entries without a key were included in the data analysis but are not visualized due to their proximity to other points.}
\label{tab:events_key_ru}
\end{table*}

\begin{table*}[t!]
\centering
\small
\begin{tabular}{m{0.1\textwidth}p{0.75\textwidth}m{0.1\textwidth}}
\textbf{Date} & \textbf{Event} & \textbf{Key} \\ 
\hline
\addlinespace[1mm] 
2022-02-08 & Putin claims allowing Ukraine to join NATO would increase the prospects of a Russia-NATO conflict that could turn nuclear. & 1b \\ 
\hline
\addlinespace[1mm] 
2022-02-21 & Putin cites Nazism in Ukraine in speech legitimizing upcoming invasion.
 & 2b \\ 
\hline
\addlinespace[1mm] 
2022-02-24 & Russia invades Ukraine.
 & - \\ 
\hline

\addlinespace[1mm] 
2022-03-02 & Russia captures Kherson.
 & - \\ 
\hline

\addlinespace[1mm] 
2022-03-21 & Russian troops used stun grenades and gunfire to disperse a rally of pro-Ukrainian protesters in the occupied southern city of Kherson on Monday.
 & 3b \\ 
\hline
\addlinespace[1mm] 
2022-03-21 & Russia abandons Kherson. & - \\ 
\hline
\addlinespace[1mm] 
2022-04-01 & Reports of Russian atrocities in Bucha begin to surface. & - \\ 
\hline
\addlinespace[1mm] 
2022-07-03 & Russia captures Lysychansk, all of Luhansk Oblast & 4b \\ 
\hline
\addlinespace[1mm] 
2022-08-29 & Ukraine launches first major counteroffensive. & 5b \\ 
\hline
\addlinespace[1mm] 
2022-09-21 & Ukraine forces Russian retreat. & 6b \\ 
\hline
\addlinespace[1mm] 
2022-11-11 & Ukraine recaptures Kherson. & 7b \\ 
\hline
\addlinespace[1mm] 
2022-12-29 & Major Russian missile attack on infrastructure facilities in Kyiv, Kharkiv, Lviv, and other cities. & 8b \\ 
\hline
\addlinespace[1mm] 
2023-02-09 & Russia launches second spring offensive. & 9b \\ 
\hline
\end{tabular}
\caption{Key events in the war discussed by Ukrainian war Bloggers. The ``Key'' column corresponds to the labeled vertical lines in Figure \ref{fig:othering_dynamics}. Entries without a key were included in the data analysis but are not visualized due to their proximity to other points.}
\label{tab:events_key_uk}
\end{table*}

\end{document}